\documentclass[aps,pra,reprint,groupedaddress,showpacs,longbibliography]{revtex4-1}

\usepackage{graphicx, xcolor}
\usepackage{times}

\newcommand{\nc}{n_{c}}\newcommand{\nqc}{n_{q}}\newcommand{\mean}{\bar n}
\usepackage[draft,color=yellow]{pdfcomment}

\usepackage{hyperref}

\begin{document}

\title{Cross-over to quasi-condensation: mean-field theories
and beyond}

\author{Carsten Henkel}
\author{Tim-O. Sauer}
\affiliation{Institute of Physics and Astronomy, University of Potsdam,
Karl-Liebknecht-Str.\ 24/25, 14476 Potsdam, Germany}

\author{N. P. Proukakis}
\affiliation{Joint Quantum Centre (JQC) Durham-Newcastle, 
School of Mathematics and Statistics, Newcastle University, 
Newcastle upon Tyne NE1 7RU, United Kingdom}

\date{11 January 2017}

\begin{abstract}
We analyze the cross-over of a 
homogeneous
one-dimensional Bose gas from the ideal gas into the
dense quasi-condensate phase. We review a number of
mean-field theories, perturbative or self-consistent,
and provide accurate evaluations of equation of
state, density fluctuations, and correlation functions.
A smooth crossover is reproduced by classical-field 
simulations based 
on the stochastic Gross-Pitaevskii equation,
and the Yang-Yang solution 
to the one-dimensional Bose gas.
\end{abstract}


\maketitle

The achievement of Bose-Einstein condensation in ultracold,
dilute atomic vapours has opened a wide research field at 
the crossroads of quantum optics and condensed-matter 
physics~\cite{PitaevskiiBook, ProukakisBook}.
It was clear from the beginning that the scenarios of the
ideal Bose gas, well-known from thermodynamics, are not sufficient
because of interactions between the atoms. In contrast 
to liquid helium where interactions are 
strong~\cite{PopovBook, GriffinBook}, ultracold vapours
can be modelled nearly from first principles, the s-wave scattering
length being the main relevant coupling constant. In lower
spatial dimensions, interactions qualitatively change the
phase diagram and lead to the emergence of 
a paired vortex phase (Kosterlitz--Thouless transition, 2D) 
or a quasi-condensate (1D).
The phase boundaries are fuzzy, however, and the cross-over region, 
as it has been called, poses a challenge to conventional pictures.
Indeed, thermal and quantum fluctuations
that are prominent anyway in lower dimensions
have to be modelled in the presence of interactions. 
As the density is lowered through the cross-over region,
density fluctuations become comparable to phase fluctuations,
so that a Luttinger liquid approach~\cite{Haldane1981b}
breaks down.
It is then questionable whether one can operate a clean splitting 
into a `quasi-condensate' and a `thermal cloud' familiar
from spontaneous symmetry breaking. 
This may explain 
why the cross-over
is so difficult to describe with mean-field theories 
that build on the Bogoliubov prescription.

In this paper, we provide a critical assessment of mean-field theories
for the description of the cross-over in a 
homogeneous
one-dimensional Bose gas
between the ideal gas and the quasi-condensate. These approaches
have the common feature that the many-body system is broken down 
to relevant collective observables that are treated as `hydrodynamic 
fields', examples 
being
the (total) density or a c-number valued 
condensate field. The hydrodynamic fields parametrise 
an approximate form of
the many-body Hamiltonian which is simple enough 
to be diagonalised in a quasi-particle basis with
a well-defined dispersion relation. This permits to compute
different mean values and correlation functions. 
Throughout the paper, we exclude 
the case of strong interactions which leads to fermionisation
(impenetrable bosons, Tonks-Girardeau regime).

There are a number
of variants for mean-field theories:
some are based on perturbative expansions (weak interactions, 
weak density fluctuations) whose validity becomes doubtful in the
cross-over region,
others are constructed in a `self-consistent' way and may suggest
a comprehensive treatment of both regimes.
We give an overview on several approaches
and work out in detail the equation of state, density 
fluctuations, and correlation functions.
There are numerous approaches that have been implemented 
to improve on the simple mean-field theories.
For a unified-notation review, the readers are referred to~Refs.\cite{Proukakis2008, ProukakisBook}.
The ``G1'' variant of the
Hartree-Fock-Bogoliubov approximation attempts to fix the issue 
of a gapless dispersion relation by carefully observing features 
a successful theory might have 
-- indeed it had some success in modelling experiments
\cite{Proukakis1998b, Hutchinson1998}.
Much has been written about a gapless spectrum in relation to
the Goldstone and
Hugenholtz-Pines theorems~\cite{Griffin1996, Hutchinson2000}. We find here
that its impact is marginal with respect to the performance
of a mean-field theory in the cross-over of the one-dimensional
Bose gas. In fact, we analyze two approximations, one gapless
(modified Popov approximation
of Refs.~\cite{Andersen2002c, AlKhawaja2002b, Proukakis2006c}),
the other not
(Hartree-Fock-Bogoliubov theory developed by
Walser and the Holland group~\cite{Walser1999, Proukakis2001, Walser2004}).
Their predictions for the equation of state
and density fluctuations are qualitatively similar, however
(see Figs.~\ref{fig:eqn-state-all4}, \ref{fig:G2-all4}). 
They also fail both in providing a smooth description 
as the chemical potential crosses zero, and predict a
critical point (discontinuity in the equation of state). 
This artefact has been noted before for
mean-field theories in three dimensions, 
see Refs.\cite{Yukalov2008a, OlivaresQuiroz2010}.
A different fate arises when the self-consistent and gapless
theory of Yukalov and Yukalova~\cite{Yukalov2008a, Yukalov2014}
is extrapolated to a one-dimensional system.
The integrals giving the non-condensate density and other
quantities diverge in the infrared, similar to the simpler
Bogoliubov theory. In Ref.\cite{Yukalov2008a},
this is claimed to be removed with dimensional regularization,
effectively subtracting the divergent piece, although the
resulting `density' becomes negative. An infrared regularization
has also been operated in the modified Popov theory, 
but following a different argument: 
the infrared-divergent pieces were identified
as spurious phase fluctuations and eliminated. The resulting
expressions are discussed here.

Another important development was the construction 
of an expansion for large particle numbers, 
but in a number-conserving way, 
following arguments laid out in Ref.~\cite{Gardiner1997b}
and extended in Refs.~\cite{Castin1998, Morgan2000, Gardiner2007, Billam2013}.
We mention that for our system of interest,
the homogeneous Bose gas in the thermodynamic limit,
the predictions of mean-field theory are qualitatively quite
similar, whether it is formulated in a number-conserving way
or in the grand-canonical ensemble with symmetry breaking.
We have checked this with the example of
extended Bogoliubov theory developed by Mora and 
Castin~\cite{Mora2003}: one key technique,
the projector orthogonal to the condensate mode,
is irrelevant for a homogeneous system where the elementary
excitations naturally appear at finite momentum.
The expansions behind these approaches, for example in the
fraction of non-condensed particles, are bound to break down
in the cross-over because there is no condensate in the dilute
phase. In the case of extended Bogoliubov theory, it is the
assumption of weak density fluctuations that fails. 

It turns out that none
of the mean-field theories analyzed here
describes the cross-over of the Bose gas
from dilute to dense in a satisfactory way: 
some theories are
simply restricted `by construction' 
to either side of the phase boundary.
Other theories give wrong predictions for one side, or suggest a
critical point, e.g., a discontinuity in the equation of state.
Fortunately, it is possible to follow the cross-over with 
the help of complex-field simulations (stochastic Gross-Pitaevskii
equation, sGP, for a review, see~\cite{Proukakis2008, Cockburn2009}).
Proposals of this technique date back to 
Stoof's group~\cite{Stoof1999a, Bijlsma2001},
Davis and the Burnett group~\cite{Davis2001c},
and Gardiner's group~\cite{Gardiner2002a, Gardiner2003}. 
See also related classical field work by 
Goral and the Rz\c{a}\.{z}ewski group~\cite{Goral2002}.
The sGP has been applied to one-dimensional Bose correlations
by one of us~\cite{Proukakis2006c}. With a suitably chosen
cutoff, its predictions are in excellent agreement with
experiments in one~\cite{Cockburn2011b}
and two dimensions~\cite{Gallucci2012}.
We find that these simulations 
successfully achieve a reasonable modelling of the entire 
cross-over.
Another `benchmark' is provided by the exact solution of the
Lieb-Liniger model at finite temperature by Yang and Yang
\cite{Lieb1963b, YangYang1969}. This approach has been used
to cross-check perturbative calculations of density fluctuations
in the dilute phase by Kheruntsyan and the Shlyapnikov group~\cite{Kheruntsyan2003, Kheruntsyan2005}.

To conclude with a comparison to experiments, it should be
noted that 
{many setups require modelling beyond the one-dimensional regime,}
mainly because the transverse confinement is not
strong enough. As the ratio between trap frequencies is
changed, one observes a ``dimensional cross-over'' from a
true three-dimensional condensate to a one-dimensional
quasi-condensate with large phase fluctuations~\cite{Petrov2001}.
Following relatively early anisotropic expansion 
experiments~\cite{Hellweg2003a, Gerbier2003},
theoretical work on this has been performed by
Al Khawaja et al.~\cite{AlKhawaja2003c} and 
Gerbier~\cite{Gerbier2004}.
Experimental work by the Bouchoule 
group~\cite{Esteve2006, Trebbia2006, Armijo2010, Armijo2011} demonstrated, for example,
the breakdown of Hartree-Fock mean field theory 
by analyzing the density fluctuations,
and mapped out the dimensional cross-over (for a review,
see Ref.\cite{BouchouleChapter2011}).
{Setups deeply in the one-dimensional regime} 
have been reported in Refs.\cite{vanAmerongen2008, Kruger2010}
where Yang-Yang thermodynamics could be checked.
The failure of mean field theories becomes 
manifest experimentally in the boundary regions of a 
trapped system~\cite{Trebbia2006}.
For the comparison with theory, 
the local density approximation (LDA) is often 
applied.
We check the accuracy of this approximation using sGP simulations
for both a homogeneous system and a trapped one.

Structure of the paper:
the problem setting and a few salient parameters are
outlined in Sec.~\ref{s:mft}. 
We discuss mean-field theories that do not
operate a splitting of the Bose gas in components 
(Sec.~\ref{s:one-component}): the ideal gas and Hartree-Fock
theory are covered~\cite{PitaevskiiBook}. The Bogoliubov approximation in Sec.~\ref{s:Bogoliubov} introduces the condensate concept,
although it suffers from serious infrared divergencies
in low dimensions. 
An extended version that applies to a quasi-condensed gas
whose density fluctuations are weak has been developed by
Mora and Castin~\cite{Mora2003}, Sec.~\ref{s:MC}.
So-called self-consistent theories are covered in
Sec.~\ref{s:self-consistent}, 
beginning with the modified Popov theory,
Sec.~\ref{s:modified-Popov}. This is based
on suitably regularised expressions for the non-quasi-condensate
component. We also illustrate in this section the 
many-body effects that renormalise the interatomic scattering
properties. The last mean-field theory is a variant of
Hartree-Fock-Bogoliubov developed by Walser,
Sec.~\ref{s:HFB-Walser}. The results are discussed in
Sec.~\ref{s:discussion} and compared to stochastic
simulations with the Gross-Pitaevskii equation. 
The Appendices summarize more technical material related
to high- and low-temperature approximations and to 
numerical aspects.

\section{Problem setting}
\label{s:mft}

\begin{table}\caption{\label{t:hydn-fields} Hydrodynamic fields.
The colons denote normal ordering of the field operators.
}
	\hrule
\begin{tabular}{rr@{${}={}$}l}
(quasi)condensate & $\phi$ 
	&
	$\langle \psi \rangle
	\,,\qquad
	\nqc = |\phi|^2$
\\
mean density & $\mean$
	&
	$\langle n \rangle = \langle \psi^\dag \psi \rangle$
\\
field correlations & $G_1( z - z' )$
	&
	$\langle \psi^\dag( z ) \psi( z' ) \rangle$
\\
density correlations & $C( z - z' )$
	&
	$\langle n( z ) n( z' ) \rangle - \mean^2$
\\
pair correlations & $G_2( z - z' )$
	&
	$\langle :n( z ) n( z' ): \rangle$
\\
thermal density & $n'$
	&
	$\mean - \nqc = G_1( 0 ) - \nqc$
\\
anomalous average & $m'$
	&
	$\langle \psi^2 \rangle - \phi^2$

\end{tabular}
\hrule
\end{table}

We consider a gas of $N$ bosonic particles of mass $M$,
strongly confined into a one-dimensional trap of length $L$,
and in thermal equilibrium at temperature $T$.
Throughout we work in the thermodynamic limit of a large
system with density $\mean = N/L$. This density is
controlled by the chemical potential $\mu$, and the 
interaction energy per particle is given by $g n$ with
a positive constant $g$. In the language of second quantization,
the Hamiltonian $H$ is
\begin{equation}
H =
\int\!{\rm d}z \left[
\frac{ \hbar^2 }{ 2M }
\frac{ d \psi^\dag }{ d z }
\frac{ d \psi }{ d z }
+
\frac{ g }{ 2 } 
\psi^\dag\psi^\dag \psi\psi
- \mu \psi^\dag \psi
\right]
\label{eq:def-Hamiltonian-density}
\end{equation}
where the field satisfies the bosonic commutation relations
$[ \psi( z ) \,, \psi^\dag( z' ) ] = \delta( z - z' )$.
A list of relevant observables is given in Table~\ref{t:hydn-fields}.
Note that for the homogeneous system we consider in this paper,
local averages like the mean density $\mean$ are spatially
constant, while correlation functions depend only on the 
distance $z - z'$ between the observation points.

The characteristic scales for the cross-over can be motivated
as follows.
For negative chemical potentials, the density is low, and
the ideal gas is a good approximation. The statistics of the
complex field operator $\psi$ is then Gaussian, and from its
fourth moment, one finds that density fluctuations
are significant: $\langle n^2 \rangle \approx 2 \mean^2$.
The cross-over is reached from below
when the interaction energy in Eq.(\ref{eq:def-Hamiltonian-density})
becomes relevant, i.e., for $\mu \sim - g \mean$. When the 
density is estimated with the degenerate ideal gas formula
[first term of Eq.(\ref{eq:ideal-gas-eqn-state}) below],
we get $\mu \sim - \mu_x$ with a characteristic energy scale
\begin{equation}
\mu_x = \left( \frac{ g M^{1/2} k_B T }{ \hbar } 
\right)^{2/3}
\label{eq:cross-over-mu}
\end{equation}
We shall see below that $\mu_x$ gives the typical width of the
cross-over region around $\mu = 0$. Repulsive interactions stabilize 
the gas so that also positive chemical potentials become accessible.
On this dense side of the cross-over, density fluctuations get
weaker: $\langle n^2 \rangle \approx \mean^2$. 
The phase still fluctuates strongly enough to prevent
the formation of long-range order, leading to the quasi-condensate
concept. 

\begin{table}[bth]
\caption{\label{t:numbers} Two sets of typical parameters
used in simulations (stochastic Gross-Pitaevskii equation).
}
\hrule
\begin{tabular}{lll}
	& Na-23		& Rb-87
\\
scattering length & $51.97\,a_0$ & $95.41\,a_0$
\\
transv. confinement	& $1.46\,{\rm kHz}$ & $4\,{\rm kHz}$
\\
interaction $g$			& $0.39\,{\rm nK\,\mu m}$	
						& $1.938\,{\rm nK\,\mu m}$
\\
temperature $T$			& $7\,{\rm nK}$	& $50\,{\rm nK}$
\\
thermal wavelength $\lambda$
		& $1.74\,\mu{\rm m}$	& $0.33\,\mu{\rm m}$
\\
$\hbar^2 k_B T / M g^2$
		& $10^3$				& $74$
\\
cross-over chem. pot. $\mu_x$
		& $0.70\,{\rm nK}$ ($14.6\,{\rm Hz}$)
		& $12\,{\rm nK}$ ($250\,{\rm Hz}$)
\\
cross-over density $n_x$
		& $1.8\,\mu{\rm m}^{-1}$ & $6.1\,\mu{\rm m}^{-1}$
\\
healing length $\xi_x$
		& $2.75\,\mu{\rm m}$ & $0.34\,\mu{\rm m}$
\end{tabular}
\hrule
\end{table}

Typical numbers are listed in Table~\ref{t:numbers} for
two different atoms~\cite{Julienne2002, WeinerBook}. 
We use the standard formula 
$g = 2\hbar\omega_\perp a_s$ for the
one-dimensional interaction constant,
assuming that the
transverse confinement gives the highest energy scale.
We define the thermal wavelength as 
$\lambda = \hbar (M k_B T)^{-1/2}$
and the healing length for a given density $n$ as
$\xi = \hbar (4 M g n )^{-1/2}$
[see Table~\ref{t:length-scales}].
In the cross-over, the two length scales are comparable,
while the density is still high enough to be far
from the Tonks-Girardeau limit. This can be expressed
in terms of the Lieb-Liniger
parameter $1/(2 n_x \xi_x)^2 \ll 1$~\cite{Lieb1963b}.

\begin{figure*}[tbh]
\includegraphics*[width = 0.9\columnwidth]{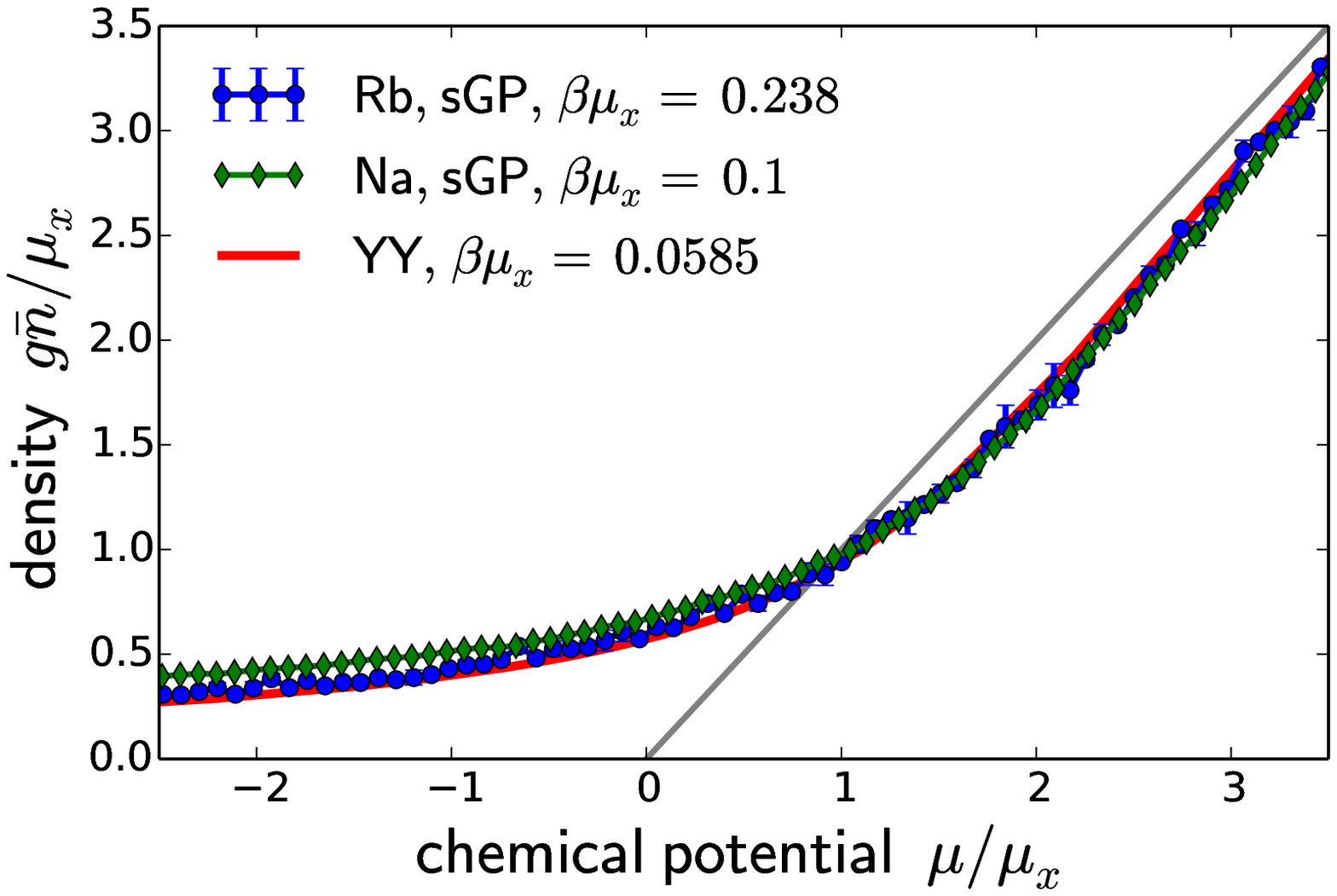}
\includegraphics*[width = 0.9\columnwidth]{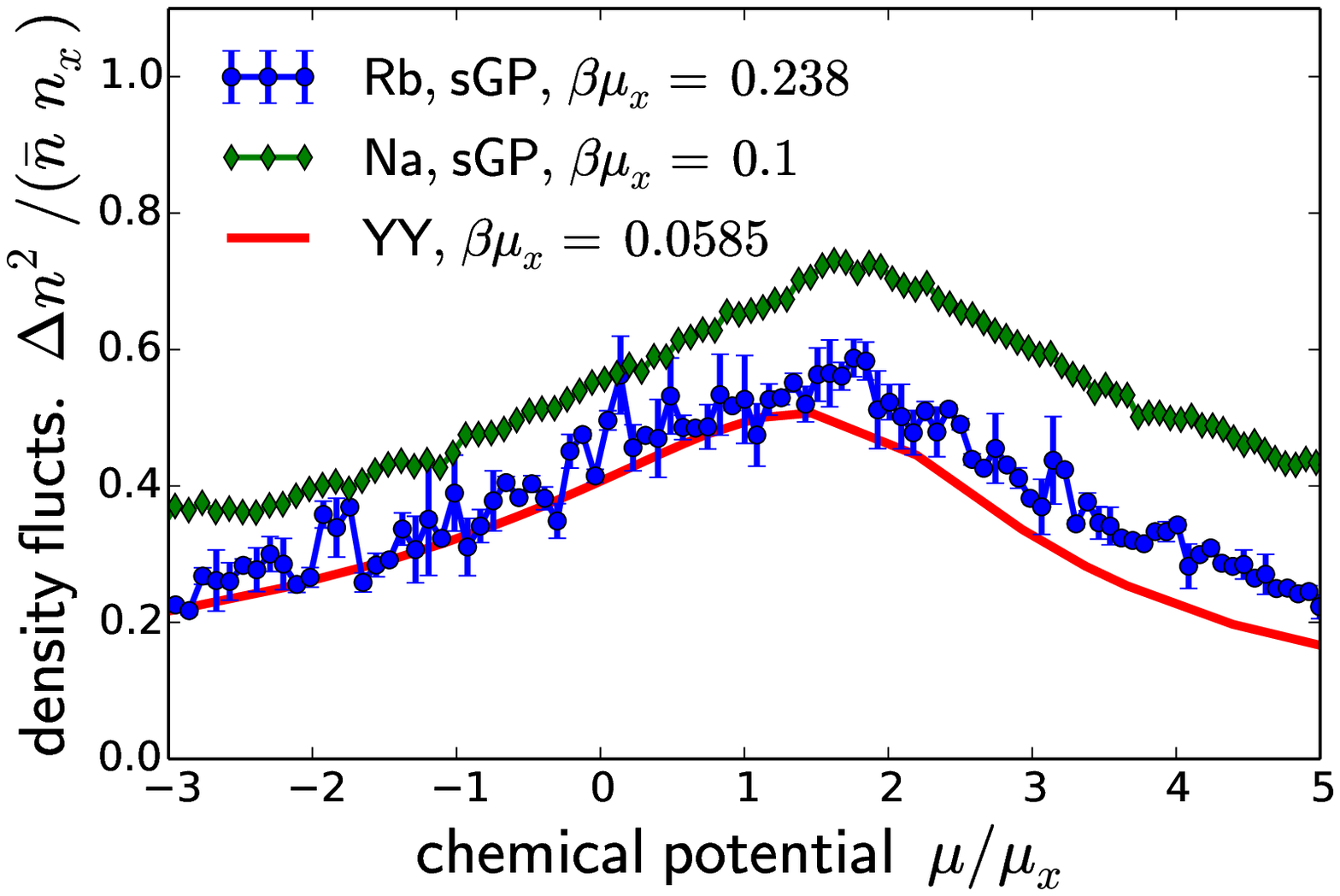}
\caption[]{Illustration of the universal features of the
dilute-to-dense cross-over for different atoms and temperatures, 
when scaled to cross-over units.
Symbols (sGP): stochastic Gross-Pitaevskii equation using the
parameters of Table~\ref{t:numbers}~\cite{Cockburn2009},
courtesy of Stuart Cockburn for the Rb data.
Solid line (YY):
numerical evaluation of the Yang-Yang solution to the
finite-temperature Lieb-Liniger model~\cite{Lieb1963b, YangYang1969},
courtesy of Karen Kheruntsyan. The temperature parameter
for the YY data is $k_B T \hbar^2 / (M g^2) = 5\times 10^3
= (\beta\mu_x)^{-3}$
[see Table~\ref{t:numbers}].
\\ 
(\emph{left}) Equation of state.
(\emph{right}) Density fluctuations, expressed by the
``Mandel parameter'' $\Delta n^2 / \mean$. (The name is chosen
by analogy
to super-Poissonian photon number distributions in laser
theory.)
}
\label{fig:simulations}
\end{figure*}

An illustration of the relevance of the energy 
scale $\mu_x$ [Eq.(\ref{eq:cross-over-mu})] is provided by
Fig.\ref{fig:simulations} where we show the equation of state
and the normalized density fluctuations for two `benchmark
theories': the first is based on numerical simulations
of the stochastic Gross-Pitaevskii equation, 
the second is the \emph{ab initio}
solution of the finite-temperature Lieb-Liniger model (Yang-Yang 
thermodynamics)~\cite{Lieb1963b, YangYang1969}. We parametrize
the data by the dimensionless inverse temperature 
$\beta\mu_x = \mu_x / k_B T$. Since this scales as
$\sim T^{-1/3}$, the temperature range covers nearly 
two orders of magnitude, but the data `collapse' into a quite
narrow band. The scale $\mu_x$ obviously captures the width 
of the cross-over zone vs.\ $\mu$ with excellent accuracy.
The density fluctuations [right panel] show a slightly larger 
scatter, but
this is due in part to a dependence on the numerical parameters
like spatial grid spacing.
{For the appropriate
choice of simulation parameters which give agreement with 
experimental data, 
see Refs.\cite{vanAmerongen2008, Armijo2010, Cockburn2011b}.
}

\begin{table*}
\caption[]{\label{t:overview-integrals}
Formulas for mean-field theories.
\\
The dispersion relations are given in the thermodynamic 
sense: the energy $\varepsilon$ appears in the Bose-Einstein 
factor $N(\varepsilon)$. 
Non-condensate density, anomalous density.
Equation of state. The lower limits of the chemical potential
are taken from Table~7.2 of Ref.\cite{SauerMSc},
$a_1 = -\zeta(\frac12)/\sqrt{2\pi} \approx 0.583$.
This is based on a low-energy (high-temperature) expansion;
higher-order corrections are ${\cal O}( \beta \mu_x )$
or ${\cal O}( \beta \mu_x )^{3/2}$.
}
\newlength{\sepl}
\setlength{\sepl}{0.9\textwidth}
\mbox{}\\[-2ex]
\begin{tabular}{lll}
Dispersion relation
\\
\multicolumn{3}{l}{\rule[2ex]{\sepl}{0.2pt}}
\\[-2ex]
Ideal gas & $\epsilon - \mu \equiv \hbar^2 k^2 / 2M - \mu$
& (gap)
\\
Hartree-Fock & $\epsilon + 2 g n' - \mu$
& (gap)
\\
Bogoliubov & $E \equiv \sqrt{\epsilon (2 g \nc + \epsilon)}$
\\
Mora-Castin & $E$ \qquad (with $g \nc \mapsto \mu$)
\\
modified Popov & $E$ \qquad (with $\nc \mapsto \nqc$)
\\
Walser & $\sqrt{(\epsilon - 2 g m') (2 g \nc + \epsilon)}$
& (gap)
\\
\multicolumn{3}{l}{\rule[2ex]{\sepl}{0.2pt}}
\\
Non-condensate density
& $\displaystyle n' = \int\!\frac{ {\rm d}k }{ 2\pi } \ldots$
& 
\\
\multicolumn{3}{l}{\rule[2ex]{\sepl}{0.2pt}}
\\[-2ex]
Ideal gas & $N( \epsilon - \mu )$
\\
Hartree-Fock & $N( \epsilon + 2 g n' - \mu )$
\\[1ex]
Bogoliubov & $\displaystyle
\frac{ \epsilon + g \nc }{ E } N( E ) 
	+ \frac{\epsilon + g \nc - E }{ 2E }$
	& (IR divergent)
\\[1ex]
Mora-Castin & $\displaystyle
\frac{ \epsilon }{ E } N( E ) 
	+ \frac{\epsilon - E }{ 2E }$ 
	& (non-positive)
\\[1ex]
modified Popov & $\displaystyle
\frac{ \epsilon }{ E } N( E ) 
	+ \frac{\epsilon - E}{2 E} + \frac{g \nqc}{2 (\epsilon + \mu)}$
\\[1ex]
Walser & $\displaystyle
\frac{ \epsilon + g(\nc - m') }{ E } N( E ) 
	+ \frac{\epsilon + g(\nc - m') - E }{ 2E }$ 
\\[1ex]

\multicolumn{3}{l}{\rule[2ex]{\sepl}{0.2pt}}
\\
Anomalous density
& $\displaystyle m' = \int\!\frac{ {\rm d}k }{ 2\pi } \ldots$
\\
\multicolumn{3}{l}{\rule[2ex]{\sepl}{0.2pt}}
\\[-2ex]
Bogoliubov & $\displaystyle
- \frac{ g \nc }{ E } \big( N( E ) + {\textstyle\frac 12} \big)$
	,
	& (IR divergent)
\\[1ex]
Walser & $\displaystyle
- \frac{ g (\nc + m') }{ E } \big( N( E ) + {\textstyle\frac 12} \big)$
\\[1ex]
\multicolumn{3}{l}{\rule[2ex]{\sepl}{0.2pt}}
\\
Equation of state
\\
\multicolumn{3}{l}{\rule[2ex]{\sepl}{0.2pt}}
\\[-2ex]
Ideal gas & $\mean = n'_{\rm id}(\mu)$
	& $\mu < 0$
\\
Hartree-Fock & $\mean = n'_{\rm id}(\mu - 2 g \mean)$
\\
Bogoliubov & $\mu = g \nc$ 
	& $\mu > 0$
\\
Mora-Castin & $\mu = g \mean + g n'$
	& $\mu / \mu_x \agt 0.630 - a_1 (\beta\mu_x)^{1/2}$
\\
modified Popov & $\mu = g \nqc + 2 g n'$
	& $\mu / \mu_x \agt 1.89 - 2 a_1 (\beta\mu_x)^{1/2}$
\\
Walser & $\mu = g \nc + 2 g n' + g m'$	
& $\mu / \mu_x \agt 2.11 - 2 a_1 (\beta\mu_x)^{1/2}$
\end{tabular}
\end{table*}

\section{One-component theories}
\label{s:one-component}

\subsection{Ideal gas}
\label{s:ideal-gas}

The simplest example is the ideal gas ($g = 0$) where the
Hamiltonian is bilinear and diagonal 
in the plane wave basis (dispersion relation 
$\epsilon( k ) = \hbar^2 k^2 / 2M$, $- \infty < k < +\infty$)
\begin{eqnarray}
H &=& \int\!{\rm d}k\,\left( \epsilon( k ) - \mu \right)
	a^\dag( k ) a( k )
\,,\nonumber
\\
\psi( z ) &=& \int\!{\rm d}k\,
{
	a(k) \frac{ \exp( {\rm i} k z ) }{ \sqrt{ 2\pi } } 
}
\label{eq:Fourier-field}
\end{eqnarray}
with annihilation and creation operators
$[ a( k ) \,, a^\dag( k' ) ] = \delta( k - k' )$. In thermal
equilibrium, we have $\langle a( k ) \rangle = 0$
and recover Bose-Einstein statistics ($1/\beta = k_BT$)
\begin{eqnarray}
\langle a^\dag( k ) a(k')\rangle &=& 
N( \epsilon( k ) - \mu ) \delta( k - k' )
\nonumber
\\
N( \epsilon( k ) - \mu ) &=&
\frac{ 1 }{ \exp \left[ \beta ( \epsilon( k ) - \mu ) \right] - 1 }
\label{eq:def-BE-average}
\end{eqnarray}
The field correlation function is denoted 
$G_1( z - z' ) = \langle \psi^\dag( z ) \psi( z' ) \rangle$
and given by
\begin{equation}
G_1( x ) = 
\int\!\frac{ {\rm d}k }{ 2\pi }
N( \epsilon( k ) - \mu )
\exp( {\rm i} k x )
\label{eq:ideal-gas-G1}
\end{equation}
The ``Boltzmann approximation'' $N( \epsilon - \mu ) \approx 
{\rm e}^{ -\beta ( \epsilon - \mu) }$ applies in the regime
$\epsilon - \mu \gg k_B T$ 
and gives a Gaussian correlation function
\begin{equation}
G_1( x ) \approx 
\frac{ {\rm e}^{ \beta \mu } }{ \sqrt{2\pi}\, \lambda }
{\rm e}^{ - x^2 / (2 \lambda^2) }
\label{eq:gaussian-limit}
\end{equation}
with a correlation length set by the thermal wavelength 
$\lambda$. If large distances are of interest or the
chemical potential approaches
the critical value $\mu = 0$, a different approximation is required.
The Rayleigh-Jeans approximation,
$N( \epsilon - \mu ) \approx k_B T / ( \epsilon - \mu )$,
captures the contribution of small $k$-modes with high 
degeneracy and yields 
an exponential shape
\begin{equation}
G_1( x ) \approx \frac{ \ell }{ \lambda^2 } 
{\rm e}^{ - |x| / \ell }
\label{eq:G1-ideal-gas}
\end{equation}
with a much larger correlation length 
$\ell = \hbar (- 2 M \mu)^{-1/2}$
[see Table~\ref{t:length-scales}].
We sketch in Appendix~\ref{a:Bose-g-half} 
an expansion 
that gives the next-to-leading order corrections
for $\lambda \ll \ell$: 
\begin{eqnarray}
\mean( \mu ) &\approx&
\frac{ \ell }{ \lambda^2 
}
- \frac{ a_1 }{ \lambda }
+ \frac{ a_2 }{ 2 } \frac{ \lambda }{ \ell^2 }
\label{eq:ideal-gas-eqn-state}
\end{eqnarray}
They are relatively important and involve the
positive coefficients 
$a_1 = - \zeta(\frac12) / \sqrt{2\pi} \approx 0.5826$
and
$a_2 = - \zeta(-\frac12) / \sqrt{2\pi} \approx 0.0830$
and the regularized zeta function.

The density correlation function [Table~\ref{t:hydn-fields}]
is computed from the Wick theorem
because the Hamiltonian $H$ [Eq.~\ref{eq:Fourier-field}] generates
Gaussian statistics. This results in 
\begin{eqnarray}
C( x ) 
&=&
\mean \, \delta( x ) + |G_1( x )|^2
\label{eq:ideal-d-fluct}
\end{eqnarray}
where the first term represents `shot noise' (it arises from
putting the field operators into normal order). The second term
is called `bunching' and increases the density fluctuations
to the level $\langle\colon n^2 {:}\rangle = 2 \mean^2$.

\begin{table}\caption{\label{t:length-scales} 
Correlation lengths
}
	\hrule
\begin{tabular}{rr@{${}={}$}l}
ideal gas ($\mu < 0$)
\\
correlation length & $\ell$
	&
	$\displaystyle
\hbar ( - 2 M \mu )^{-1/2}
\approx \mean \lambda^2$
\\[1ex]
Bogoliubov theory ($\mu > 0$)
\\
phase diffusion length\footnote{Over the distance 
$\ell_\theta$, the phase quadrature 
$Y$~[Eq.(\ref{eq:integral-phase-Brown})] has diffused
such that its variance is comparable to the condensate
density $\nc = \mu/g$.}
	& $\ell_\theta$
	&
	$\nc \lambda^2$
\\
healing length & $\xi$
	&
		$\hbar (4 M \mu )^{-1/2}$
\\[1ex]
extended Bogoliubov ($\mu > 0$)
\\
phase correlation length
	& $\ell_\theta$
	& $2 \mean \lambda^2$
\\
density correlation length
	& $\xi$
	& $\hbar (4 M \mu )^{-1/2}$
\\[1ex]
modified Popov
\\
phase correlation length
	& $\ell_\theta$
	& $2 \nqc \lambda^2$
\\
density correlation lengths
	& $\xi_q$
	& $\hbar (4 M g \nqc)^{-1/2}$
	\\
	& \multicolumn{2}{l}{and $\sqrt{2}\, \xi$}
\\[1ex]
Hartree-Fock-Bogoliubov
\\
field and density correlation lengths
	& $\xi_c$
	& $\hbar (4 M g \nc)^{-1/2}$
\\
and	& $\xi_m $
    & $\hbar (- 4 M g m')^{-1/2}$
\end{tabular}
\hrule
\end{table}

\subsection{Interacting gas: Hartree-Fock}
\label{s:Hartree-Fock}

\begin{figure*}[tbh]
\centerline{\includegraphics*[height = 0.6\columnwidth]{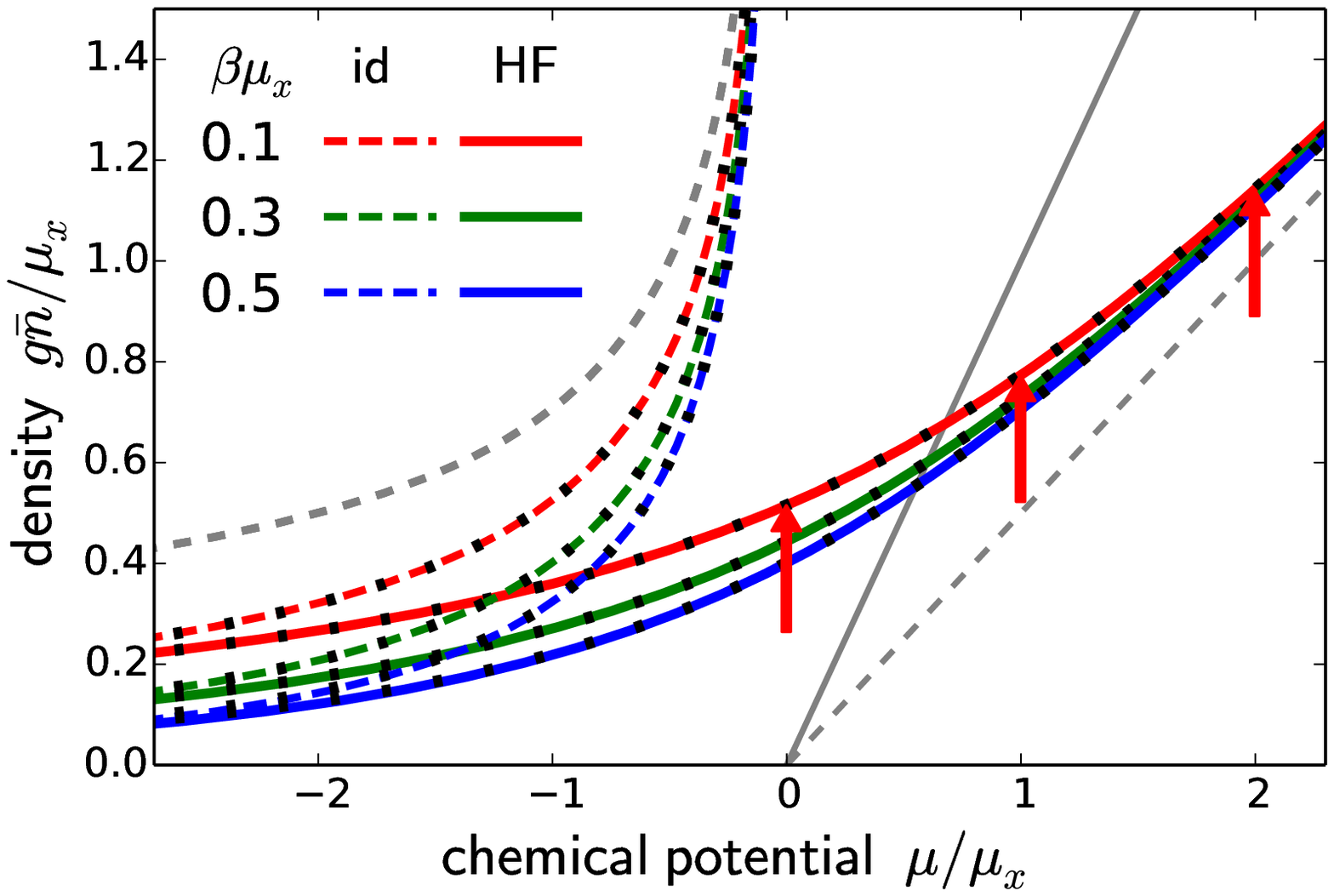}
\hspace*{05mm}
\includegraphics*[height = 0.6\columnwidth]{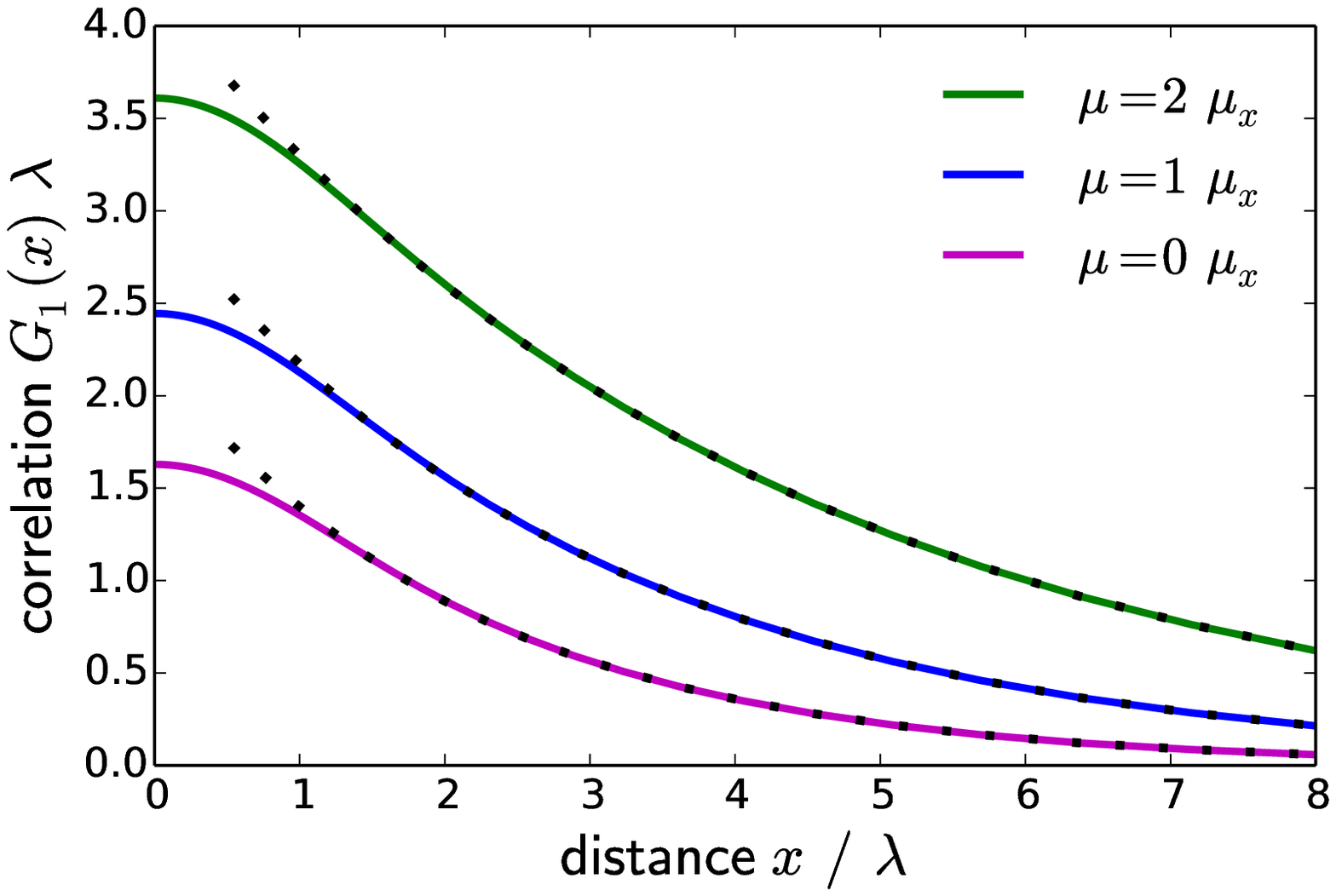}
}
\caption[]{Comparison of one-component mean-field theories.
\\
(\emph{left}) equation of state $\mean(\mu)$. Dashed: ideal
gas, solid: Hartree-Fock theory. Dashed gray curve: classical
approximation (first term of Eq.(\ref{eq:ideal-gas-eqn-state})).
Dotted lines, superimposed: low-energy approximations based
on Eq.(\ref{eq:ideal-gas-eqn-state}).
Straight solid line: pure condensate $\mu = g \mean$,
straight dashed: Hartree-Fock asymptote
$\mu = 2 g \mean$. Chemical potential and density scaled
to cross-over units [Eq.(\ref{eq:cross-over-mu})]. Arrows:
values of $\mu$ selected for right panel.
\\
(\emph{right}) correlation function $G_1( x )$ for three
different densities (marked by red arrows on the left), 
Hartree-Fock theory. Temperature
such that $\beta\mu_x = 0.1$. Dotted curves: low-energy
approximation based on Eq.(\ref{eq:G1-ideal-gas}); their
characteristic length (decay to $1/{\rm e}$) is
$\ell \approx \mean \lambda^2$.
The same results would have been obtained for an ideal
gas at the chemical potentials (from top to bottom)
$\mu \approx -0.285, -0.547, -1.03\,\mu_x$
[see Table~\ref{t:overview-integrals}].
}
\label{fig:id-HF}
\end{figure*}

Hartree-Fock theory is probably the oldest mean-field theory;
it is treating the interactions in a Bose gas in terms of
an additional potential (the `mean field'). The
Hamiltonian is approximated in the plane-wave basis
by
\begin{equation}
H \approx 
\int\!{\rm d}k \left(
\epsilon( k ) + 2 g \mean - \mu \right)
a^\dag( k ) a( k )
\label{eq:HF-Hamiltonian}
\end{equation}
This shift of the chemical potential gives the same equation
of motion as the full interaction Hamiltonian when correlation
functions are factorized in a 
Gaussian approximation~\cite{Griffin1996}.
As long as $2 g \mean > \mu$, Bose-Einstein statistics can be
applied as for the ideal gas, and we get the following implicit
equation for the (mean) density
\begin{equation}
\mean = 
\int\!\frac{ {\rm d}k }{ 2\pi }
\frac{ 1 }{ 
\exp \left[ \beta (\epsilon( k ) + 2 g \mean - \mu) \right] 
- 1 }
\label{eq:HF-density}
\end{equation}
To work out this formula,
we use an ideal-gas chemical potential
$\mu_i < 0$ as parameter and plot $\mean_{\rm id}( \mu_i )$
vs.\ $\mu = \mu_i + 2 g \mean_{\rm id}( \mu_i )$.
The approximation shown in
Eq.(\ref{eq:ideal-gas-eqn-state}) can also be used here; 
it is fairly accurate,
as long as $\beta \mu_x \alt 0.1$ (see dotted lines in
Fig.\ref{fig:id-HF}(\emph{left})).

In the leading order, 
we get the explicit expression
\begin{equation}
\mu \approx 2 g \mean - \frac{ k_B T }{ 2 (\mean \lambda)^2 }
= 2 g \left( \mean - \frac{ n_x^3 }{ 4 \mean^2 } \right)
\label{eq:HF-density-expanded}
\end{equation}
Right at the cross-over
$\mu = 0$, we have $\mean = 2^{-2/3}\,n_x \approx 0.63\,n_x$ 
where the cross-over density scale $n_x = \mu_x / g$ is 
defined by Eq.(\ref{eq:cross-over-mu}).
In the dense case ($\mu \gg \mu_x$), note again the collapse 
of the data in cross-over units over a wide range of temperatures.
The equation of state
$\mu \approx 2 g \mean$, however, is off by
50\% compared to Bogoliubov theory
[Eq.(\ref{eq:Bogo-eqn-state}) below, 
see Fig.\ref{fig:id-HF}(\emph{left})]. 
This will be improved
by a more advanced mean-field theory.

The correlation function $G_1(x)$ of Hartree-Fock theory
is formally given
by the same integral~(\ref{eq:ideal-gas-G1}) as for the ideal gas,
but evaluated at the self-consistent chemical potential,
as shown in Fig.\ref{fig:id-HF}.
The density correlations
are given by Eq.(\ref{eq:ideal-d-fluct}) because
of Gaussian statistics. They therefore show 
the same bunching as the ideal Bose gas.
At large distances, they feature an exponential decay 
on a length scale
$\ell \approx \mean \lambda^2$ 
[see Eqs.(\ref{eq:G1-ideal-gas}, \ref{eq:ideal-gas-eqn-state})]
that is much 
larger than the thermal wavelength. 
This behaviour does not capture the strong differences
between phase and density fluctuations that characterise
the dense phase~\cite{Haldane1981b}. 
(An improved version of Hartree-Fock including
the many-body renormalization of particle interactions is
briefly discussed in Sec.~\ref{s:HF-m}.)

\section{Expansion around a (quasi) condensate}

\subsection{Bogoliubov theory}
\label{s:Bogoliubov}

\begin{table}\caption{\label{t:Bogo-theory} 
Bogoliubov quasi-particles
($\mu = g |\phi|^2 > 0$).}
	\hrule
\begin{tabular}{rr@{${}={}$}l}
free particle energy & $\epsilon(k)$
	&
	$ \hbar^2 k^2 / 2M $
\\
dispersion relation & $E(k)$
	&
	$ \sqrt{ \epsilon( k ) (2 \mu + \epsilon( k ) ) }$
\\[1ex]
Bogoliubov amplitudes & $u(k)$
	&
	${\rm e}^{ {\rm i} \varphi }\, \cosh \frac12\alpha(k)$
\\[1ex]
	(condensate phase $\varphi$) & $v(k)$
	&
	$-{\rm e}^{ {\rm i} \varphi }\, \sinh \frac12\alpha(k)$
\\[0.5ex]
&$\cosh \alpha(k)$&
	$\displaystyle
	\frac{ \epsilon(k) + \mu }{ E(k) }$
\\[0.5ex]
&$\sinh \alpha(k)$&
	$\displaystyle
	\frac{ \mu }{ E(k) }$
\vspace*{0.5ex}
\end{tabular}
\hrule
\end{table}

This mean field approach is very successful in three dimensions
and implements the concept of spontaneous symmetry breaking
in the dense phase. Although it is not directly applicable
in lower dimensions, it provides an introduction to the key
concepts. The basic idea is the so-called Bogoliubov shift
where the field operator is split into a c-number valued field
(the `condensate') and fluctuations,
$\psi \mapsto \phi + \hat\psi$. 
The Hamiltonian is
expanded up to second order in the fluctuations, and 
the condensate is determined by the requirement that the terms 
linear in $\hat\psi$ vanish. This gives the
Gross-Pitaevski equation
\begin{equation}
- \frac{ \hbar^2 }{ 2 M } \frac{ d^2\phi }{ dz^2 }
+ g |\phi(z)|^2 \phi = \mu \phi
\label{eq:stationary-GPe}
\end{equation}
For reasons of thermodynamic stability, one chooses the
condensate with the largest possible density---which is
spatially constant in a homogeneous system.
We get the equation of state
\begin{equation}
\mu = g |\phi|^2 = g \nc
\label{eq:Bogo-eqn-state}
\end{equation}
that leaves the phase of $\phi$ undetermined. 
The conventional choice of
real and positive $\phi$ can be interpreted as a spontaneous
breaking of the U(1)-symmetry of the field 
Hamiltonian~(\ref{eq:def-Hamiltonian-density}).
The self-interaction of the condensate contributes an energy density
$\epsilon_c = -\frac12 g \nc^2$ to the Hamiltonian,
corresponding to a pressure
$p = g \nc^2 / 2$. These parameters allow for acoustic elementary excitations
with a speed of sound $c$ at long wavelengths set by
$M c^2 = \partial p / \partial \nc = g \nc$.

\subsubsection{Quasi-particle spectrum}

The part of the Hamiltonian that is of second order in 
$\hat\psi$ is diagonalized with the help of the
Bogoliubov transformation
\begin{equation}
\hat\psi( z ) = \int\!
\frac{ {\rm d}k }{ \sqrt{ 2\pi } }
\left[
b( k ) u( k ) \, {\rm e}^{ {\rm i} k z} + b^\dag( k ) v( k )\, {\rm e}^{ -{\rm i} k z} \right]
\label{eq:Bogoliubov-trafo}
\end{equation}
where the Bogoliubov amplitudes $u$ and $v$ are given
in Table~\ref{t:Bogo-theory}. They are constrained by
$|u|^2 - |v|^2 = 1$ to make
the operators $b$ and $b^\dag$ bosonic
[commutation relation
$[ b(k) , \, b^\dag(k') ] = \delta( k - k' )$,
as after Eq.(\ref{eq:Fourier-field})].
We note that both $u$ and $v$ are proportional
to the phase factor ${\rm e}^{ {\rm i} \varphi }$ involving
the condensate phase. 
The operator $b(k)$ annihilates a quasi-particle
with energy $E(k)$ given by the
Bogoliubov dispersion relation~[Table~\ref{t:Bogo-theory}]
where the acoustic branch involves the speed of sound 
$c = (g \nc/M)^{1/2}$ consistent with the hydrodynamic 
argument mentioned above.
In this long wavelength limit, 
the Bogoliubov amplitudes become comparable and large,
$u \sim - v \gg 1$.

Finally, the field Hamiltonian is truncated at second order 
in $\hat\psi$, taking the following form,
\begin{equation}
H \approx \epsilon_0 L +
\int\!{\rm d}k \, E(k) b^\dag(k) b(k) 
\label{eq:diagonal-Hamiltonian}
\end{equation}
The zero-point energy density $\epsilon_0$ arises by
putting the Bogoliubov operators $b$, $b^\dag$ into normal order.
It is given by the integral
\begin{equation}
\epsilon_0 - \epsilon_c 
=
-
\int\!\frac{ {\rm d}k }{ 2\pi } 
E(k) |v(k)|^2
= - 
\int\!\frac{ {\rm d}k }{ 2\pi } 
\frac{ \mu^2 / 2 }{
E(k) + \epsilon(k) + \mu
}
\label{eq:Bogo-zero-pt-energy}
\end{equation}
which converges and can be computed analytically
(Appendix~\ref{a:zero-T-expansion})
\begin{equation}
\epsilon_0 
= 
- \frac{ \mu \nc }{ 2 } 
- \frac{ \mu }{ 3\pi \xi }
.
\label{eq:Bogo-gd-energy}
\end{equation}
The Bogoliubov correction is small and scales with the
Lieb-Liniger parameter~\cite{Lieb1963b}
$1 / (\nc \xi)$
where $\xi$ is the healing length of 
Table~\ref{t:length-scales}. 
More analytical results at zero temperature in one and higher 
dimensions can be found in Refs.\cite{Lee2003,Astrakharchik2006}.

The key problem of Bogoliubov theory in one dimension is
the infrared divergence of the non-condensate density $n'$. The
latter is defined as
\begin{equation}
n' = \langle \psi^\dag \psi \rangle - |\phi|^2 = 
\langle \hat\psi^\dag \, \hat\psi \rangle
\label{eq:}
\end{equation}
In thermal equilibrium with respect to the approximate 
Hamiltonian~(\ref{eq:diagonal-Hamiltonian}),
the modes corresponding to the $b(k)$'s
have an occupation $N(E(k))$ [see Eq.(\ref{eq:def-BE-average})].
Using the expansion~(\ref{eq:Bogoliubov-trafo}) of the field
operator, the thermal density is given by the integral
\begin{eqnarray}
n' &=& \int\!\frac{ {\rm d}k }{ 2\pi } n'( k )
\\
n'(k) &=&
N(E(k)) |u(k)|^2
+
[ N(E(k)) + 1 ] |v(k)|^2
\label{eq:thermal-density}
\end{eqnarray}
The zero temperature limit $N( E ) \to 0$ gives the so-called 
depletion density that arises by the scattering of virtual
particles out of the condensate.
Its integral is divergent in the infrared (IR)
because the 
Bogoliubov amplitude scales $|v(k)|^2 \sim 1/k$ at long wavelengths. 
The temperature-dependent part shows an even stronger divergence 
$\sim T / k^2$ so that Bogoliubov theory is only useful
as a conceptual framework. 
Here is an explicit expression 
for the non-condensate distribution in $k$-space 
that will re-surface later [see also Table~\ref{t:overview-integrals}]
\begin{equation}
n'(k) = 
\frac{ \mu + \epsilon( k ) }{ 2 E(k) }
\coth\frac{ \beta E(k) }{ 2 }
- \frac12 
\label{eq:Bogo-thermal-d}
\end{equation}
For completeness, we mention that in (symmetry-broken)
Bogoliubov theory, also the so-called anomalous average 
$m = \langle \psi \psi \rangle = \phi^2 + m'$ is nonzero. 
In $k$-space, it involves the product of the two 
Bogoliubov amplitudes, 
\begin{eqnarray}
m'(k) &=& 
\left[ 2 N(E(k)) + 1 \right]
u(k) v(k)
\\
&=&
-
\frac{ \mu }{ 2 E( k ) }
\coth\frac{ \beta E(k) }{ 2 }
\label{eq:Bogo-anomalous}
\end{eqnarray}
but its integral is also IR-divergent. Note that
we fixed the condensate phase to $\varphi = 0$ in the second
line.

\subsubsection{Density and phase diffusion}

\begin{figure*}[hbt]
\centerline{\includegraphics*[width=0.6\columnwidth]{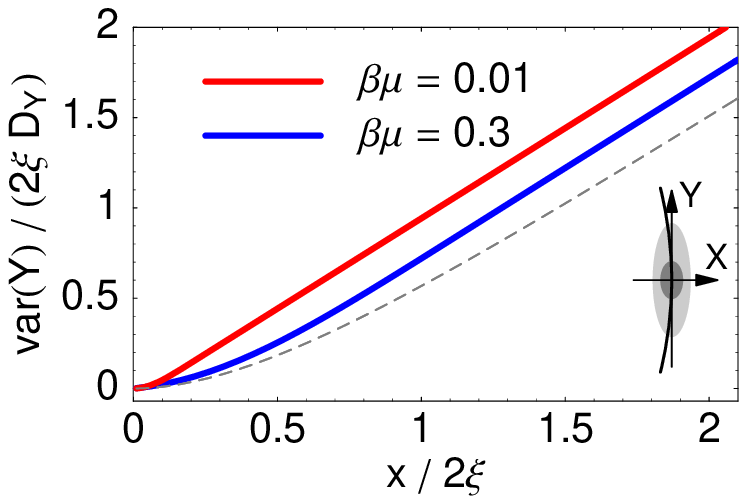}\hspace*{05mm}\includegraphics*[width=0.6\columnwidth]{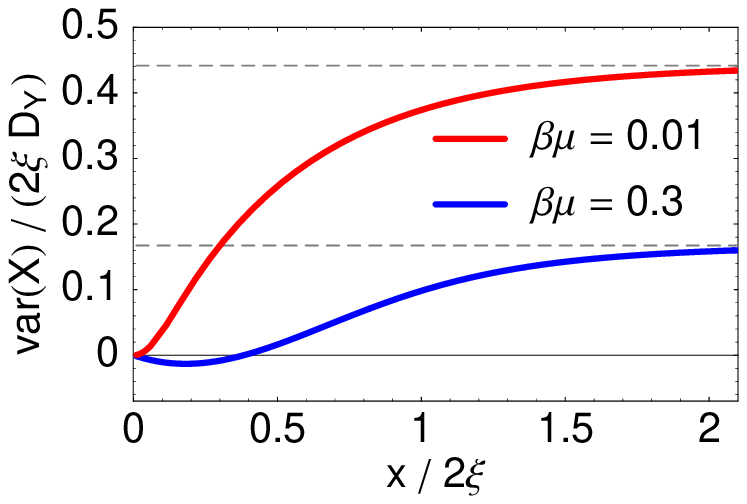}}
\caption[]{Spatial diffusion of quadrature components in
Bogoliubov theory.
We plot the variances of the imaginary (left) and real (right)
parts of the field difference $\psi( z ) - \psi( z' )
= X + {\rm i} Y$ vs.\ the
distance $x = z - z'$.
Red: low density, blue: high density,
dashed gray: analytical approximation. The variances are 
calculated in normal order with respect to the $\psi$ and
$\psi^\dag$ field operators and divided by the 
temperature-dependent `diffusion
constant' $D_Y = M k_B T / \hbar^2$ to fit onto the same scale;
the distance is scaled to the healing length $\xi$.
\\
Left: phase quadrature $\langle\colon Y^2 {:}\rangle$. 
Blue: $\mu = 0.3\,k_B T$, red $\mu = 0.01\,k_B T$.
The dashed line arises from the
low-energy expansion~(\ref{eq:smooth-diffusion}).
The inset illustrates the anisotropic diffusion of the complex
field in the `Mexican hat potential', starting from a 
symmetry-broken condensate value.
\\
Right: density quadrature  $\langle\colon X^2 {:}\rangle$. 
Blue: $\mu = 0.3\,k_B T$, red: $\mu = 0.01\,k_B T$.
The dashed lines give the 
low-energy approximation~(\ref{eq:Bogo-th-density-2})
to the large-distance plateau. Negative values correspond
to squeezing below the shot-noise level.}
\label{fig:XY-Bogo}
\end{figure*}

Finite results within Bogoliubov theory
can be produced by considering spatial increments,
similar to Brownian motion.
Consider the difference
$\Delta\psi( z, z' ) = \psi( z ) - \psi( z' )$ 
and its real and imaginary parts, assuming real and positive
$\phi$. 
The average vanishes, 
$\langle \Delta\psi \rangle 
= \langle X + {\rm i} Y \rangle = 0$,
and for the (co)variances, we find 
$\langle\colon X Y + Y X {:}\rangle = 0$
and the integral representations
\begin{eqnarray}
\langle\colon X^2 {:}\rangle &=& 
\int\!\frac{ {\rm d}k }{ 2\pi }
\left( 1 - \cos k x \right)
\left\{
\frac{ \epsilon }{ 2 E }
\coth\frac{ \beta E }{ 2 }
- \frac12
\right\}
\label{eq:integral-density-Brown}
\label{eq:integral-phase-Brown}
\\
\langle\colon Y^2 {:}\rangle &=& 
\int\!\frac{ {\rm d}k }{ 2\pi }
\left( 1 - \cos k x \right)
       \left\{
\frac{ \epsilon + 2 \mu }{ 2 E }
\coth\frac{ \beta E }{ 2 }
- \frac12
\right\}
\nonumber
\end{eqnarray}
where $x = z - z'$
and the arguments of $\epsilon(k)$ and $E(k)$ have
been dropped for brevity. The colons denote normal
ordering with respect to the 
field operators $\hat\psi$ and $\hat\psi^\dag$
(\emph{not} with respect
to the quasi-particle operators $b(k)$, $b^\dag(k)$).
By virtue of the identity
\begin{equation}
\int\!\frac{ {\rm d}k }{ 2\pi }
\frac{ 1 - \cos k x }{ k^2 } = 
\frac{ | x | }{ 2 }
\label{eq:delta-function-identity}
\end{equation}
an infrared ($k \to 0$) divergence of the integrand
translates into a `diffusive spreading' of the quantum field
$\Delta \psi$
as the distance $x$ between two positions increases.
This behaviour appears only in the phase 
quadrature (imaginary part $Y$) which asymptotes to
$\langle\colon Y^2 {:}\rangle \sim D_Y |x - x'|$.
The `diffusion constant' is given by the simple and
universal expression 
$D_Y = M k_B T / \hbar^2 = \lambda^{-2}$. This is illustrated in Fig.\ref{fig:XY-Bogo}(left) 
where the dashed line is calculated from the low-energy
approximation \begin{equation}
\int\!\frac{ {\rm d}k }{ 2\pi }
(1 - \cos k x)
\frac{ 2 \mu }{ \beta E(k)^2 } = 
D_Y\!
\left\{
| x | - 
\xi 
(1 - {\rm e}^{ - | x | / \xi } )
\right\}
\label{eq:smooth-diffusion}
\end{equation}
with $\xi$ the healing length [Table~\ref{t:length-scales}].

The density quadrature (real part $X$) 
does not diffuse freely: its variance
reaches a finite limit given by the integral in 
Eq.(\ref{eq:integral-density-Brown}) with the cosine dropped.
In Appendix~\ref{a:Bose-g-half}, we find in
the low-energy limit 
the result
\begin{equation}
|x| \gg \xi \gg \lambda:\qquad
\langle\colon X^2 {:}\rangle 
\approx 
\frac{ \xi }{ \lambda^2 } 
- \frac{ a_1 }{ \lambda }
- \frac{ a_2 }{ 4 } \frac{ \lambda }{ \xi^2 } 
\label{eq:Bogo-th-density-2}
\end{equation}
Note the close analogy of the sub-leading terms 
with the ideal gas expansion~(\ref{eq:ideal-gas-eqn-state})
where the same positive coefficients 
$a_1$, $a_2$
appear. As shown in 
Fig.\ref{fig:XY-Bogo}(right), this agrees well with the
(numerically computed) variance $\langle\colon X^2 {:}\rangle$
at large distances. Note the negative values at low
temperature (`below shot noise'): 
by analogy to quadrature fluctuations in quantum 
optics \cite{WallsMilburnBook,VogelWelschBook}, 
this can be interpreted as the squeezing of the density quadrature 
due to the nonlinear interaction with the condensate.
At zero temperature, the squeezing reaches the level
\begin{equation}
T = 0, |x| \gg \xi:\quad
\langle\colon X^2 {:}\rangle 
\approx 
- \frac{ 1 }{ 2 \pi \xi }
\,,
\label{eq:Bogo-th-density-0}
\end{equation}
as an elementary integration shows
[see Eq.(\ref{a:q-depletion-1})].

\subsubsection{Density correlations}

We finally quote the density correlations
$C( z - z' )$ [see Table~\ref{t:hydn-fields}]. Due
to the Bogoliubov shift, the density operator takes
the form $\hat n( z ) = |\phi|^2 
+ \phi^* \hat\psi( z ) 
+ \hat\psi^\dag( z ) \phi
+ \hat\psi^\dag( z ) \hat\psi( z )$, and 
we get additional contributions compared to the ideal
gas [Eq.(\ref{eq:ideal-d-fluct})]. The second- and fourth-order
correlations of the fluctuation $\hat\psi$ are worked
out with the Wick theorem.
The result can be written in the form
\begin{eqnarray}
C( z - z' ) &=&
\mean \,\delta( z - z' ) 
\nonumber
\\
&& {}
 + 
2 \mathop{\rm Re} \left\{
|\phi|^2 \langle \hat\psi^\dag(z) \hat\psi(z') \rangle
+ \phi^{*2} \langle \hat\psi(z) \hat\psi(z') \rangle
\right\}
\nonumber
\\
&& {}
+ |\langle \hat\psi^\dag(z) \hat\psi(z') \rangle|^2
+ |\langle \hat\psi(z) \hat\psi(z') \rangle|^2
\label{eq:Bogo-d-fluct}
\end{eqnarray}
This formula is only partially meaningful
because the last two terms
are both infrared-divergent.
The curly bracket can be combined into a regular integral
\begin{eqnarray}
&&
2 \mathop{\rm Re} \left\{
|\phi|^2 \langle \hat\psi^\dag(z) \hat\psi(z') \rangle
+ \phi^{*2} \langle \hat\psi(z) \hat\psi(z') \rangle
\right\}
\nonumber\\
&& = 2 |\phi|^2
\int\!\frac{ {\rm d}k }{ 2\pi }
\cos k x
\left\{
\frac{ \epsilon }{ 2 E }
\coth\frac{ \beta E }{ 2 }
- \frac12
\right\}
\label{eq:convergent-G2-Bogo}
\end{eqnarray}
This is essentially the same as the `density quadrature'
$\langle\colon X^2 {:}\rangle$ 
[Eq.(\ref{eq:integral-density-Brown})]. 
Fig.\ref{fig:XY-Bogo}(\emph{right}) with a flip in orientation
can thus be interpreted as a plot of the density correlation 
function. Note the density correlation length $\xi$
that emerges from the typical $k$-scale $\epsilon(k) \sim \mu$
of the integrand.

The divergences of Bogoliubov theory
have been addressed, of course, by the other mean-field theories 
we analyze now.

\subsection{Extended Bogoliubov theory}
\label{s:MC}

Mora and Castin~\cite{Mora2003}
have based this theory on an alternative expansion
in the dense regime $\mu > 0$, using the assumption 
that phase gradients and density fluctuations are small. 
There
is no spontaneous symmetry breaking here, but rather a (phase-fluctuating) quasi-condensate. 
{For earlier work in this spirit in trapped systems, see
Ref.\cite{Petrov2000}.
}
The dilute side of the cross-over ($\mu < 0$) 
with significant bunching 
is excluded by construction,
and one should expect that the expansion breaks down as the
density is lowered.

\subsubsection{Density and phase operators}

The theory introduces a quasi-condensate component with 
density $\nqc = \mu/g$ 
[denoted $\rho_0$ in Ref.\cite{Mora2003}]
and mutually
conjugate phase and density fluctuation operators
$\hat\theta( z )$, $\delta \hat n( z )$~\cite{Haldane1981b, Petrov2000}.
The existence
of the phase operator is secured by working on a discrete
lattice and assuming the probability of zero particles
per lattice cell to be negligible.
The Hamiltonian
expanded to second order in the fluctuations can be
diagonalized and yields again the 
Bogoliubov dispersion relation $E(k)$ [Table~\ref{t:overview-integrals}]
where
$\mu/g$ appears in lieu of the condensate density $\nc$.
With this proviso, the Bogoliubov amplitudes 
$u( k )$, $v(k)$ for the fluctuation operators 
have the same structure 
as in Bogoliubov theory~[Table~\ref{t:Bogo-theory}].
The mode expansions of the fluctuation operators are
($k$-arguments suppressed for simplicity)
\begin{eqnarray}
\hat\theta( z ) &=&
\hat\varphi +
\frac{ \nqc^{-1/2} }{ 2{\rm i} }
\!\!\int\!\frac{ {\rm d}k }{ \sqrt{ 2\pi } }
\left\{
(u - v) b \, {\rm e}^{ {\rm i} k z }
- {\rm h.c.}
\right\}
\nonumber
\label{eq:phase-expansion}
\\
\delta \hat n( z ) &=&
\nqc^{1/2}
\!\!\int\!\frac{ {\rm d}k }{ \sqrt{ 2\pi } }
\left\{
(u + v) b \, {\rm e}^{ {\rm i} k z }
+ {\rm h.c.}
\right\}
\label{eq:density-expansion}
\end{eqnarray}
where $\hat\varphi$ is the operator 
for the quasi-condensate phase (spatially constant).
We have taken the thermodynamic limit where $z$ is continuous
and the momentum
conjugate to $\hat\varphi$ can be neglected.

\subsubsection{Equation of state}

\begin{figure*}[hbt]
\centerline{\includegraphics*[width = 0.9\columnwidth]{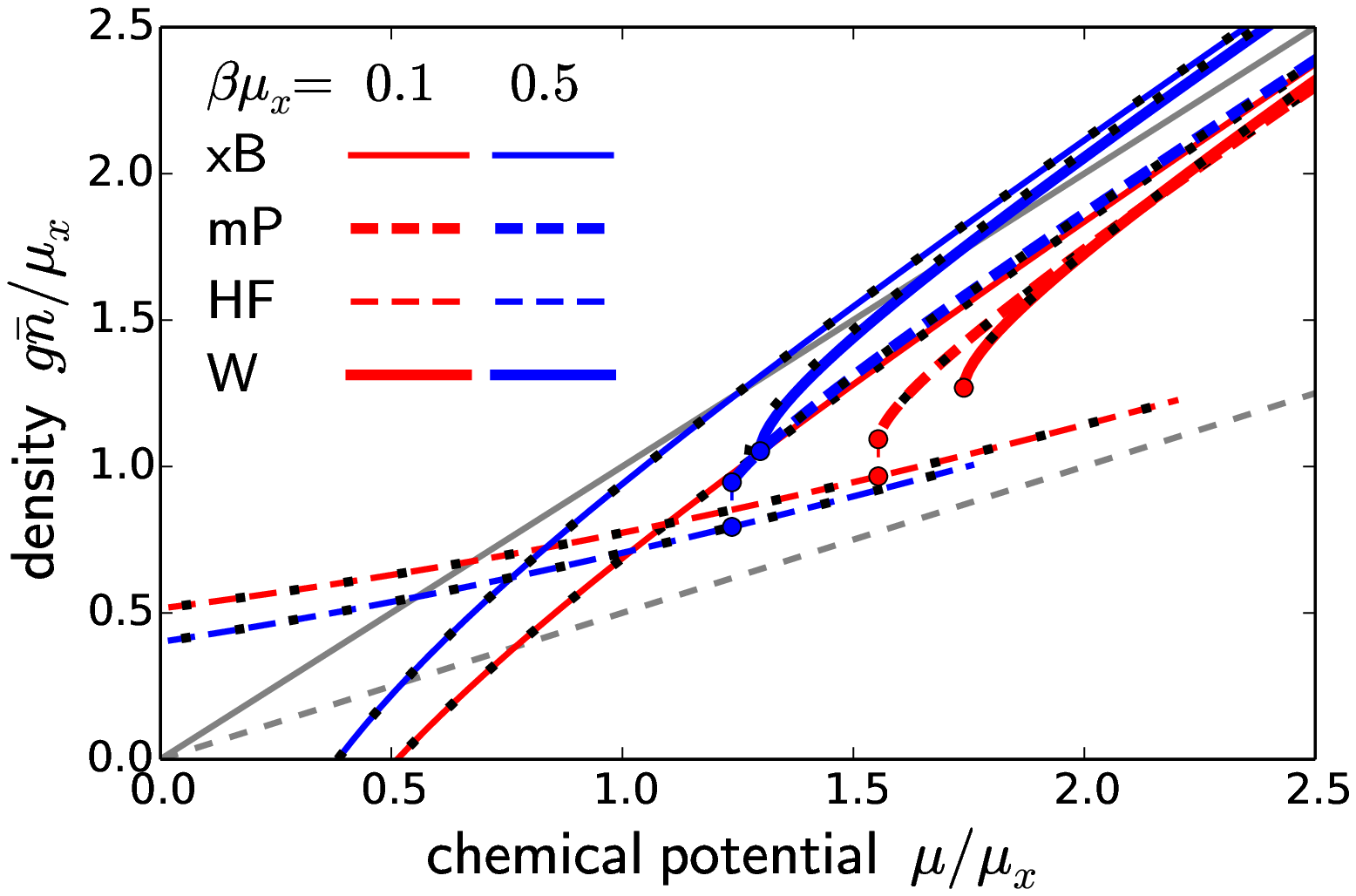}\hspace*{0.1\columnwidth}
\includegraphics*[width = 0.9\columnwidth]{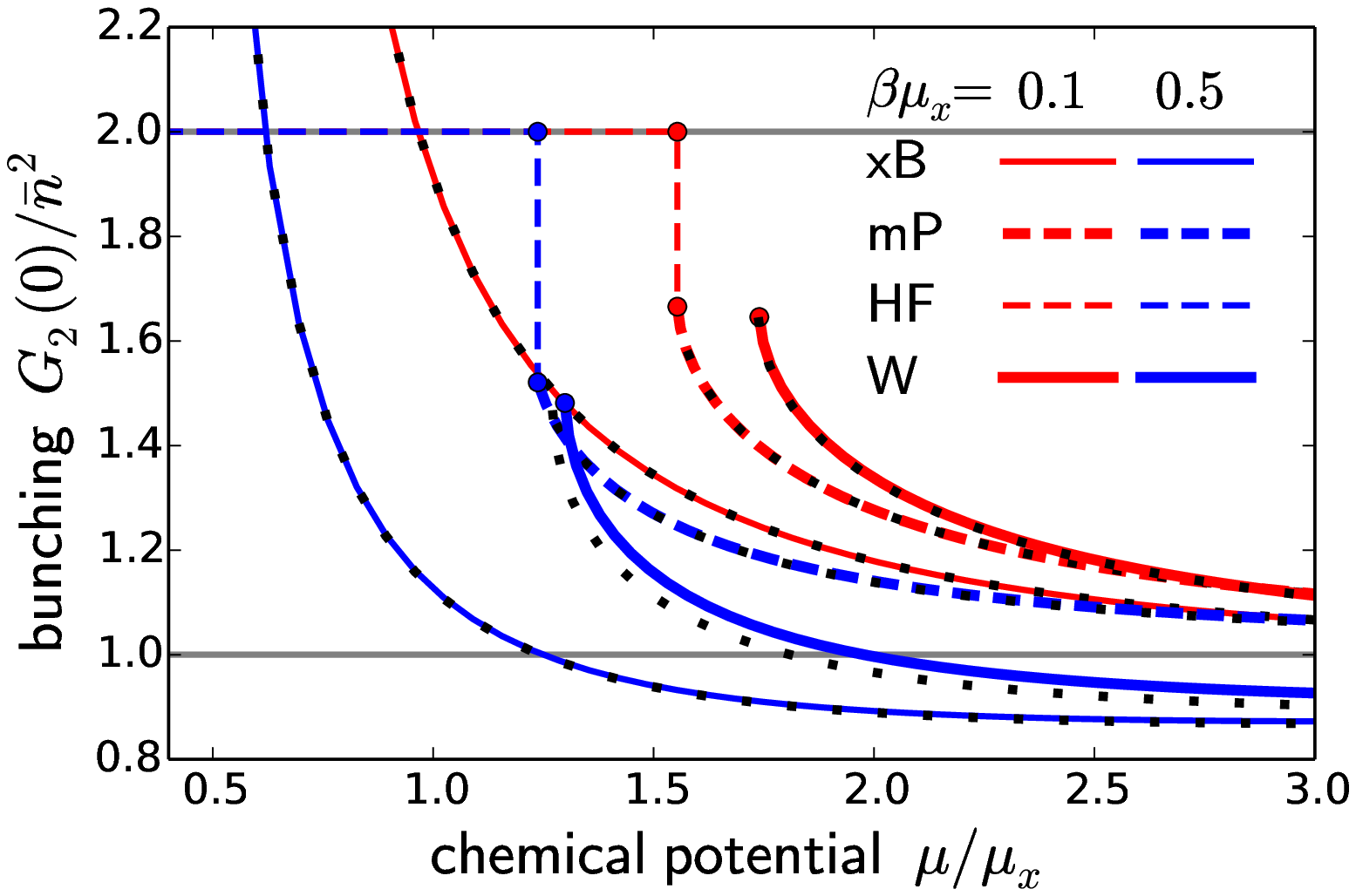}
}
\caption[]{Equation of state (\emph{left}) and density fluctuations
(\emph{right}). 
Comparison of different mean-field theories:
thin solid lines (xB) -- extended Bogoliubov theory 
(Mora and Castin, Ref.\cite{Mora2003});
dashed lines (HF, mP) -- modified Popov theory
(Andersen, Al-Khawaja et al., Ref.\cite{Andersen2002c, AlKhawaja2002b});
thick solid lines (W) -- Hartree-Fock-Bogoliubov theory
(Walser, Ref.\cite{Walser1999, Walser2004}).
Cross-over units [see Eq.(\ref{eq:cross-over-mu}) for $\mu_x$].
Dotted (superimposed) lines: low-energy approximations.
\\
(\emph{left}) Equation of state.
The upper gray diagonal corresponds to a pure condensate
(Bogoliubov theory). Lower set of dashed curves: Hartree-Fock
theory with asymptote (dashed) $2 g \mean \approx \mu$.
\\
(\emph{right}) Density fluctuations, expressed via the
pair correlation function $G_2(0)$.
Upper line $G_2(0) = 2\mean^2$: ideal gas and Hartree-Fock
approximation.
Lower line $G_2(0) = \mean^2$: pure condensate. 
Anti-bunching ($G_2(0) < \mean^2$) occurs in extended 
Bogoliubov theory for $\beta\mu \agt 0.6$.
}
\label{fig:eqn-state-MC-mP-W}
\end{figure*}

The average non-quasi-condensate density 
vanishes when computed with respect
to the second-order Hamiltonian (subscript $2$),
$\langle \delta \hat n \rangle_2 = 0$.
Third-order terms in the expansion are 
needed to describe the non-condensate density
and are taken into account in perturbation theory.
The resulting equation of state involves
the same integrand as the
density quadrature $X$ in Eq.(\ref{eq:integral-density-Brown}):
\begin{eqnarray}
\mu &=& 
g \mean + g n'
\label{eq:mu-MC}
\\
n' &=& 
\int\!\frac{ {\rm d}k }{ 2\pi }
\left\{
(u + v)^2 N( E )
+
v (u + v)
\right\}
\nonumber\\
&=&
\int\!\frac{ {\rm d}k }{ 2\pi }
\left\{
\frac{ \epsilon }{ 2 E } \coth\frac{ \beta E }{ 2 }
- \frac12
\right\}
\label{eq:nprime-MC}
\end{eqnarray}
This formula can be used to compute the mean density $\mean = \mean( \mu )$. Its structure is the same
as in modified Popov theory [Eq.(\ref{eq:mu-mP})], and we
therefore used the notation $n'$. 
Mora-Castin theory does not pretend, however, that 
$n'$ can be interpreted as a non-quasi-condensate density.
Indeed, the integral~(\ref{eq:nprime-MC}) 
becomes negative at low temperatures. At zero temperature, 
we get 
[by the same calculation as in Eq.(\ref{eq:Bogo-th-density-0})] 
\begin{equation}
T = 0:\qquad
\mean = 
\frac{ \mu }{ g } + \frac{ 1 }{ 2\pi \xi } 
\label{eq:MC-zero-T-density}
\end{equation}
with the healing length $\xi = \hbar(4 M \mu)^{-1/2}$.
Since the first term is the quasi-condensate density
$\nqc$, the second one can be interpreted as the depletion density.
In the opposite limit of high temperatures (low energies), 
we can use Eq.(\ref{eq:Bogo-th-density-2}) to get
\begin{eqnarray}
\mean &\approx &
\frac{ \mu }{ g } - 
\left\{
	\frac{ \xi }{ \lambda^2 } 
	- \frac{ a_1 }{ \lambda } 
	- \frac{ a_2 }{ 4 } \frac{ \lambda }{ \xi^2 }
\right\}
\label{eq:MC-eqn-state-low-E}
\end{eqnarray}
which is in excellent agreement with the data plotted
in Fig.\ref{fig:eqn-state-MC-mP-W}. 
One notes that the density exceeds the linear
approximation $\mean \approx \mu / g$ (light gray) 
in the dense phase,
this is due to the curly bracket in 
Eq.(\ref{eq:MC-eqn-state-low-E}) becoming
negative. We use cross-over units in this plot
[see around Eq.(\ref{eq:cross-over-mu})] and emphasize
that despite the factor 125 in temperature, the scatter
of the data is relatively small, also among the mean-field
theories.
Mora-Castin theory fails to predict a positive total density
in the cross-over region: 
for $\mu \alt 2^{-2/3}\,\mu_x \approx 0.630\,\mu_x$
[see Table~\ref{t:overview-integrals}].
This could have been expected
because in this range, density fluctuations become so large
that the expansion around a `quiet' quasi-condensate breaks
down. The size of the density fluctuations can be appreciated
from the correlation functions 
in Figs.\ref{fig:eqn-state-MC-mP-W}(\emph{right})
and~\ref{fig:g1G2-correlations}(\emph{bottom}).

\subsubsection{Correlation functions}

The field correlation 
function is found as follows [Eq.(146) of Ref.\cite{Mora2003}]
\begin{equation}
G_1( x ) = 
\mean
\exp\left[ 
- {\textstyle\frac12} 
\langle\colon \Delta \hat\theta(x)^2 {:}\rangle_2
- 
{
\frac{1}{8 \nqc^2}
}\langle\colon \Delta \delta \hat n(x)^2
{:}\rangle_2
\right]
\label{eq:Mora-G1}
\end{equation}
where the difference operators $\Delta \hat A(x) = 
\hat A(x) - \hat A(0)$ are
similar to the $\Delta \psi$ operator introduced around
Eqs.(\ref{eq:integral-density-Brown}). 
The normal-order prescription $\colon \ldots {:}$
is with respect to the fluctuation operators $\hat\psi$.
This expression includes in a perturbative way contributions
to the Hamiltonian that are of third order in the fluctuations.
The exponent in Eq.(\ref{eq:Mora-G1})
has the convergent integral representation
[Eq.(184) of Ref.\cite{Mora2003}]
\begin{eqnarray}
\log\frac{ G_1( x ) }{ \mean }
&=& 
- \frac{1}{\mean} \int\!\frac{ {\rm d}k }{ 2\pi }
(1 - \cos kx ) 
\nonumber\\
&& \quad
\left\{
\frac{ \mu + \epsilon }{ 2 E }
\coth\frac{ \beta E }{ 2 }
- \frac12 
\right\}
\label{eq:G1-exponent-MC}
\end{eqnarray}
Mora and Castin~\cite{Mora2003} have 
recognized the integrand as
the non-condensate spectrum of Bogoliubov theory
[Eq.(\ref{eq:Bogo-thermal-d})]. The infrared divergence of
the latter therefore yields an exponential decay at large
distance $x$:
$G_1( x ) \sim \exp( - |x| / \ell_\theta )$
[using Eq.~(\ref{eq:smooth-diffusion})].
The (phase) correlation length 
$\ell_\theta = 2 \mean \lambda^2$ 
is twice as large as for the ideal Bose gas 
(parameter $\ell$ in Eq.(\ref{eq:G1-ideal-gas})).
A comparison to other mean-field theories is provided
in Fig.\ref{fig:g1G2-correlations}(\emph{top}, \emph{center}).

At zero temperature, the integral~(\ref{eq:G1-exponent-MC})
diverges only logarithmically.
As explained in Appendix~\ref{a:zero-T-expansion}, 
one gets for large $x$ (here, $\gamma \approx 0.577$)
\begin{equation}
T = 0:\qquad
\log\frac{ G_1( x ) }{ \mean }
\approx
- \frac{ \log( 2 |x| / \xi ) + \gamma - 2 }{ 4\pi \mean \xi }
\,.
\label{eq:MC-zero-T-G1}
\end{equation}
The exponent of this power law has been given earlier
by Refs.\cite{Haldane1981b,Andersen2002c}, but even the prefactor
agrees with Ref.\cite{Astrakharchik2006} in the 
regime $\mean \xi \gg 1$.

For later comparison with the modified Popov theory
[Sec.~\ref{s:modified-Popov}], we also quote the formula for
phase diffusion. Keeping only terms up to second order,
one gets indeed
the phase quadrature 
$\langle\colon Y^2( x ) {:}\rangle$
of Bogoliubov theory~[Eq.(\ref{eq:integral-phase-Brown})]
\begin{eqnarray}
&&
\langle\colon \Delta \hat\theta(x)^2 {:}\rangle_2
=
\nonumber
\\
&&
\frac{ 1 }{ \nqc }
\int\!\frac{ {\rm d}k }{ 2\pi }
(1 - \cos k x) 
\left\{
\frac{\epsilon + 2 \mu}{ 2 E }
\coth \frac{ \beta E }{ 2 }
- \frac{1}{2}
\right\}
\label{eq:MC-phase-fluctuations-normal}
\end{eqnarray}
This term is at the origin of phase diffusion 
$\langle\colon \Delta \hat\theta(x)^2 {:}\rangle_2 \approx |x| 
/ (\nqc \lambda^2)$ in the exponent of $G_1(x)$.

The density correlations are obtained directly from the expansion~(\ref{eq:density-expansion}) 
of the fluctuation operator $\delta \hat n( z )$
[Eq.(121) of Ref.\cite{Mora2003}]
\begin{eqnarray}
C( z - z' ) &=& \mean \delta( z - z' )
+ 
\langle\colon \delta \hat n(z) \delta \hat n(z') {:}\rangle_2
\nonumber
\\
&\approx&
\mean \delta( z - z' ) + 2 \mean n'( x )
\label{eq:MC-density-correlations}
\end{eqnarray}
Here, $n'( x )$ is given by Eq.(\ref{eq:nprime-MC})
with an additional $\cos k x$ under the integral. 
Note that we recover the same expression as the regular
part of 
Eq.(\ref{eq:convergent-G2-Bogo}) in Bogoliubov theory.
The density correlation length is therefore of the order 
of the healing length $\xi$, much shorter than 
the characteristic phase correlation length $\ell_\theta$ 
[see after Eq.(\ref{eq:G1-exponent-MC})]. A low-energy
approximation to Eq.(\ref{eq:MC-density-correlations})
can be found by keeping only the classical part
$\coth(\beta E/2) \approx 2 / (\beta E)$ of the integrand,
leading to
\begin{equation}
x \gg \lambda: \qquad
C( x ) \approx
2 \frac{ \mean \xi }{ \lambda^2 } {\rm e}^{ - |x| / \xi }
\label{eq:MC-G2-approximation}
\end{equation}
See Fig.\ref{fig:g1G2-correlations}(\emph{bottom}) for a comparison. 
For $\mu \alt \mu_x$, density fluctuations are clearly
too large for the expansion behind Mora-Castin theory to be valid.
The squeezing of the density quadrature manifests itself 
by the non-monotonous behaviour of the pair correlation function
$G_2( x)$
as $x$ increases from zero. At zero temperature, the density 
shows some `anti-bunching'
\begin{equation}
T = 0:\qquad
G_2( 0 ) = \mean^2 - \frac{ \mean }{ \pi \xi }
< \mean^2
\,,
\label{eq:MC-anti-bunching-T=0}
\end{equation}
but this small reduction is of course far from the 
`correlation hole' of Fermi liquids 
or the Tonks-Girardeau gas~\cite{ProukakisBook,PinesNozieresBook}.

\section{Self-consistent theories}
\label{s:self-consistent}

{These theories construct a
simplified form for the Hamiltonian involving hydrodynamic 
fields.} These are fixed at a later stage
by equating them to thermodynamic averages computed 
with this approximate Hamiltonian (`self-consistency'). The 
simplest example of such a theory is Hartree-Fock
[Sec.~\ref{s:Hartree-Fock}] that works with a single
field, the density $\mean$. More elaborate methods
also include a (quasi)condensate or, for example,
the anomalous average, and aim at describing the Bose
gas also at higher densities. We discuss here two examples
in detail.

\subsection{Modified Popov theory}
\label{s:modified-Popov}

\begin{figure}[htbp]
   \includegraphics[width=0.9\columnwidth]{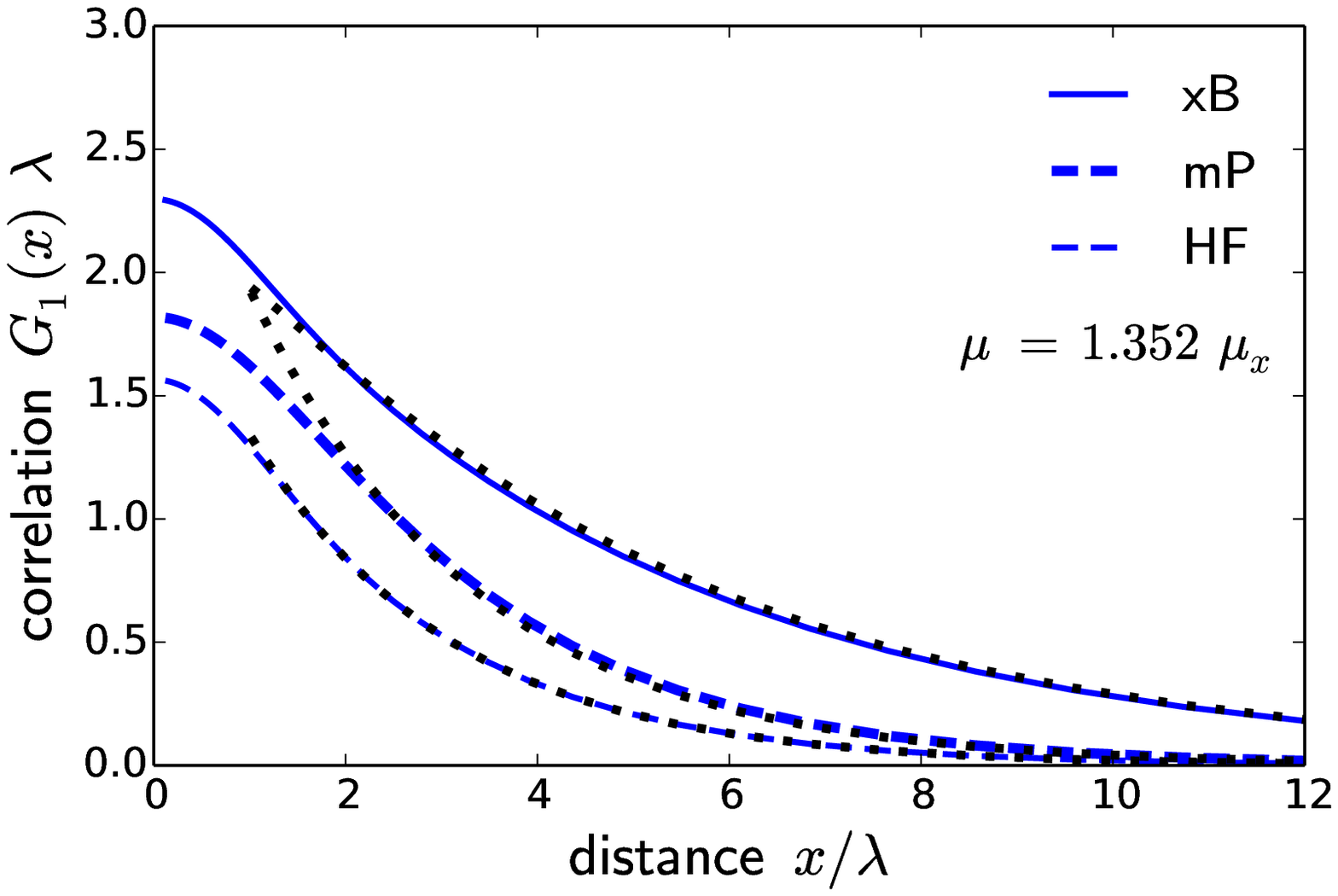}
   \includegraphics[width=0.9\columnwidth]{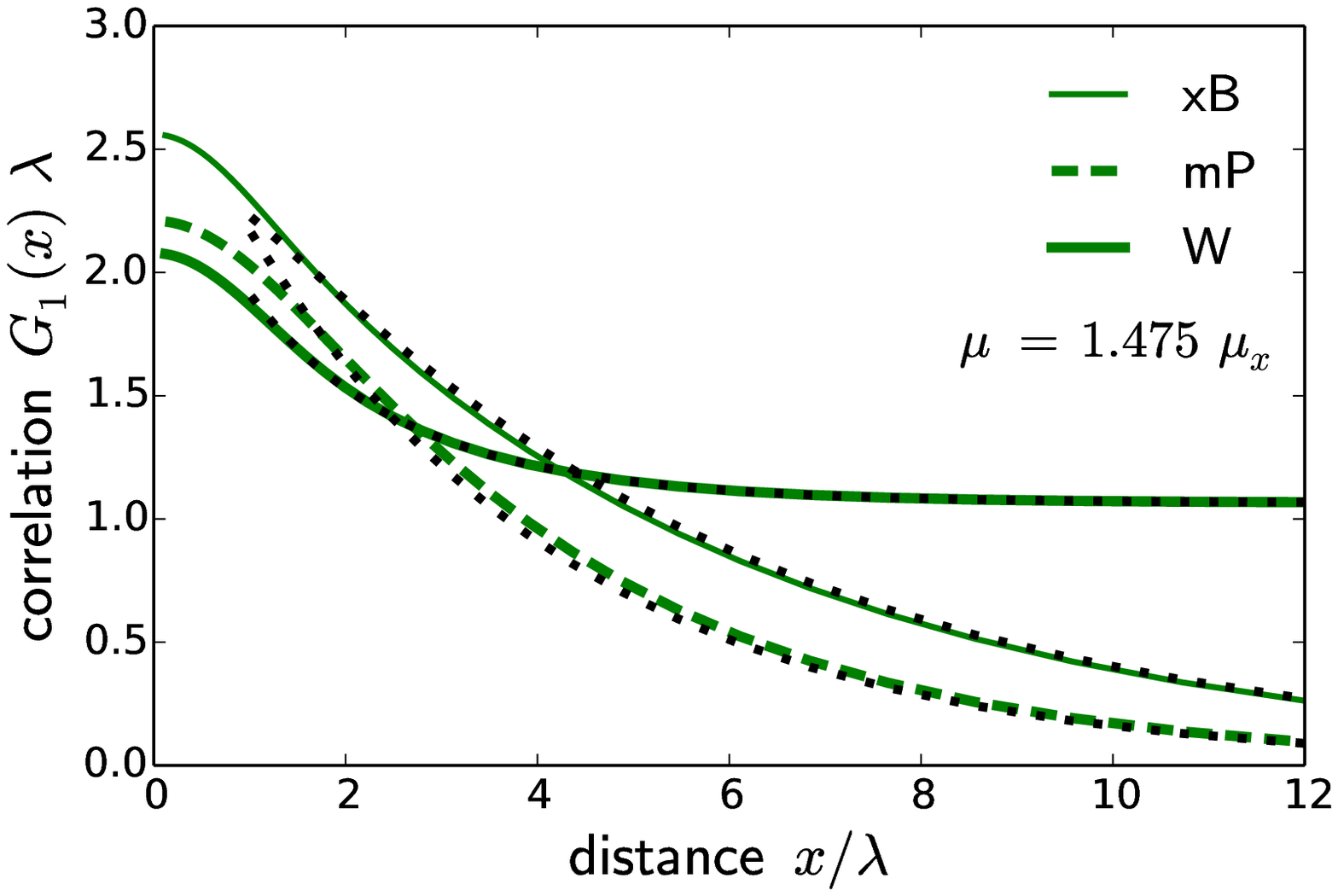}
   \includegraphics*[width = 0.9\columnwidth]{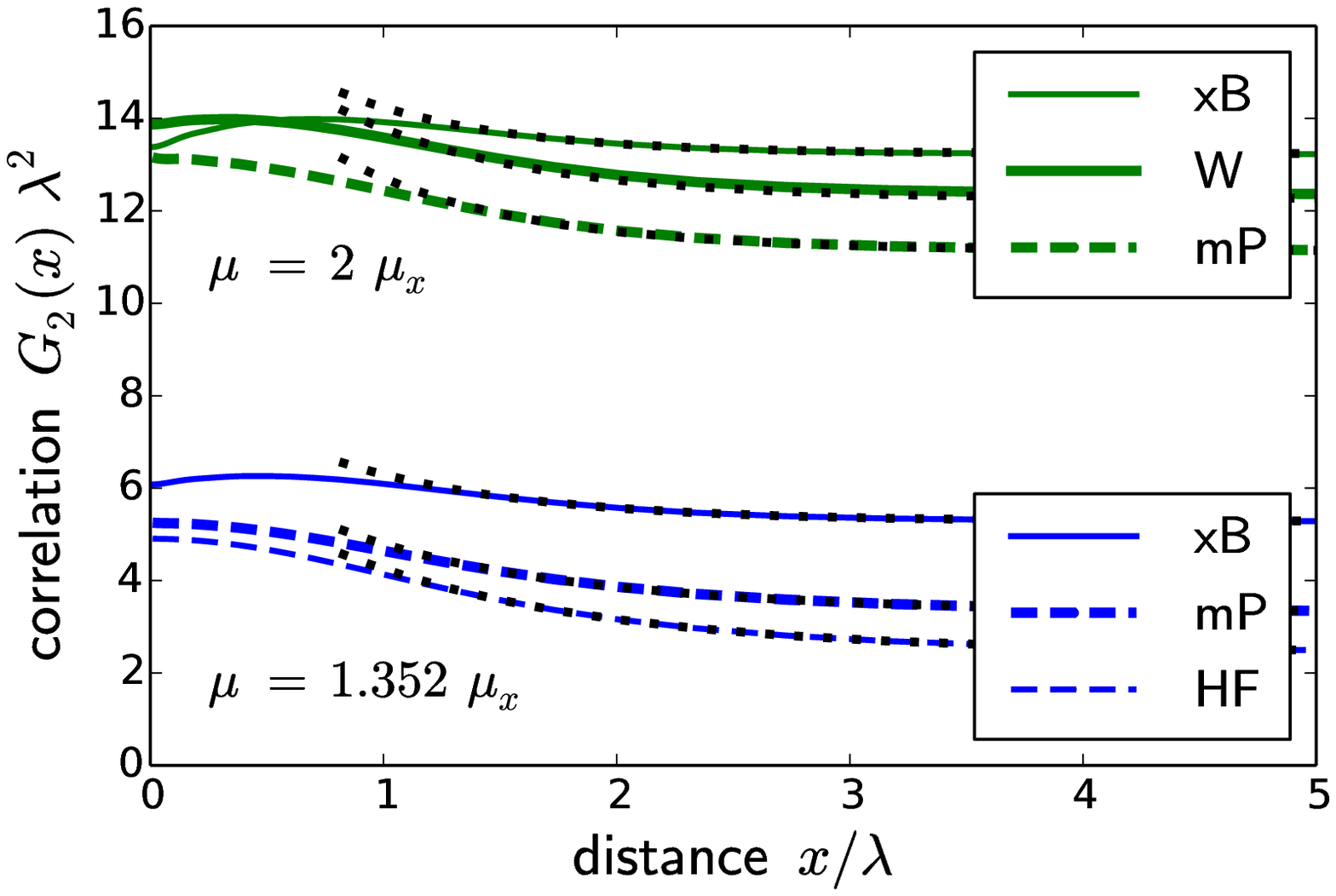}
   \caption[]{Comparison of correlation functions
between mean field theories (labelling as in
Fig.\ref{fig:eqn-state-MC-mP-W}).
\\
(\emph{top} and \emph{center})
Field correlation function $G_1( x )$
in the cross-over. Dotted curves: low-energy approximations.
The distance is scaled to the thermal wavelength. 
Temperature such that $\beta\mu_x = 0.3$.
(\emph{top}) Critical point of modified Popov theory 
where the description `jumps' between 
`HF' (dilute)
and 
`mP' (dense).
(\emph{center}) Critical
point of 
{
Hartree-Fock-Bogoliubov theory (`W' = Walser). 
}
Note the different behaviour with respect to long-range order 
(`true vs quasi-condensate').
\\
(\emph{bottom})
Pair correlation function $G_2( x )$ for two chemical
potentials. Note the smaller range of distances $x$.
Dotted lines: exponential approximations
Eqs.(\ref{eq:MC-G2-approximation}, \ref{eq:mP-G2-approximation},
\ref{eq:approx-G1-Walser}, \ref{eq:approx-mp-Walser}).
Same temperature: $\beta\mu_x = 0.3$.
Lower set of curves: critical point of modified Popov theory (mP/HF),
below the range of Hartree-Fock-Bogoliubov theory (W).
}
   \label{fig:g1G2-correlations}
\end{figure}

This mean-field theory is based on the idea that low-energy
fluctuations actually destroy the long-range order, and there
is no condensate in the ordinary sense (long-range order
\`a la Penrose-Onsager \cite{PitaevskiiBook}). 
{For details of the theory and similar approaches, 
we refer to Refs.~\cite{Andersen2002c,AlKhawaja2002b,Proukakis2006c,Petrov2000}.
}
Note that in dimensions $2$ and $3$, 
the (`bare') interaction constant $g$ 
gets renormalized into an energy- (and momentum-) dependent 
T-matrix~\cite{AlKhawaja2002b,DeSilva2014}. 
This effect is usually neglected in one-dimensional systems.
We provide a brief discussion in Sec.~\ref{s:HF-m}.
The theory
is applied differently on the two sides of the cross-over:
on the dilute side, the Hartree-Fock 
approximation is applied~[Sec.~\ref{s:Hartree-Fock}],
while the dense case is outlined now.

\subsubsection{Equation of state}

As the density increases beyond $\sim n_x$, 
the density $\mean = \nqc + n'$ 
of the system is split into the quasi-condensate
$\nqc$ and the thermal part $n'$. The former determines the
speed of sound in the (gapless) dispersion relation
\begin{equation}
E( k ) = [ 2 g \nqc \epsilon( k ) + \epsilon^2( k ) ]^{1/2}
\label{eq:modPopov-Ek}
\end{equation}
The thermal density is given by the convergent integral
\begin{eqnarray}
n'
&=& 
\int\!\frac{ {\rm d}k }{ 2\pi }
\left\{
\frac{ \epsilon }{ 2 E }
\coth\frac{ \beta E }{ 2 }
- \frac{ 1 }{ 2 }
+ 
\frac{ g \nqc }{ 2 (\epsilon + \mu) }
\right\}
\label{eq:mod-Popov-nprime}
\end{eqnarray}
where the first two terms have the same structure as
Eq.(\ref{eq:nprime-MC}).
The last term
has been introduced as a counterterm to regularize
the zero-temperature (depletion) density.
It has the merit of making Eq.(\ref{eq:mod-Popov-nprime}) 
positive
at all values of $\mu$
so that an interpretation as the density 
of the non-quasi-condensate is applicable.
The equation of state is written
\begin{equation}
\mu = g \mean + g n'
\label{eq:mu-mP}
\end{equation}
It looks formally like
Eq.(\ref{eq:mu-MC}) of extended Bogoliubov theory,
although the interpretation of the non-condensate density
$n'$ is different.
Eq.(\ref{eq:mu-mP}) is an implicit equation for 
the chemical potential, since $\mu$ also appears in $n'$.
See Appendix~\ref{a:numerics}
for details on the numerical procedure.

The zero-temperature analysis can be done similar to 
Eq.(\ref{eq:MC-zero-T-density}), and $n'$ then describes the
quasi-condensate depletion:
\begin{equation}
T = 0:\qquad
n' \approx
\frac{ \pi / \sqrt{8} - 1 }{ 2\pi \xi }
\label{eq:mP-eqn-state-T=0}
\end{equation}
where the approximation $\xi \approx \xi_q$ was used \footnote{There is probably a misprint in
Ref.\cite{Andersen2002c} and in Eq.(26) of 
Ref.\cite{AlKhawaja2002b} where the denominator is
$4\pi \xi$.}.
While the scaling with the healing length 
is the same, the prefactor differs 
from Eq.(\ref{eq:MC-zero-T-density})
due to the counterterm in Eq.(\ref{eq:mod-Popov-nprime}).
At high temperatures, the techniques of 
Appendix~\ref{a:high-T-expansion} can be used to derive
the approximation
\begin{eqnarray}
n' 
& \approx &  
\frac{ \xi_q }{ \lambda^2 }
- \frac{ a_1 }{ \lambda }
+ \frac{ \xi }{ 4\sqrt{2}\, \xi_q^2 }
- \frac{ a_2 \lambda }{ 4 \xi_q^2 }
\label{eq:mP-nc-thermal}
\end{eqnarray}
where $\xi_q = \hbar(4 M g \nqc)^{-1/2}$ 
is the healing length of the quasi-condensate density $\nqc$.

In Fig.\ref{fig:eqn-state-MC-mP-W}(\emph{left}),
the equation of state (thick dashed)
is compared to extended Bogoliubov theory (thin solid). In the
dense phase, the difference is small, the self-consistent
theory predicts a slightly lower density. In the cross-over 
region $\mu \approx 1.89\,\mu_x$, a `critical point' is reached
[see Table~\ref{t:overview-integrals}]:
below this value, the implicit equation of state has no solution. There is a finite gap to the density given by Hartree-Fock theory
(lower lines), which is the appropriate mean-field description
on the dilute side~\cite{AlKhawaja2002b}.

\subsubsection{Correlation functions}

{The first-order correlation function 
can be found, e.g., in Eq.(8) of Ref.\cite{Proukakis2006c}
}
\begin{equation}
G_1( x ) = \mean \, 
\exp[ - {\textstyle\frac12} 
\langle \Delta \theta^2( x ) \rangle_{\rm mP}
] \,,
\label{eq:G1-modPopov}
\end{equation}
it involves phase fluctuations given by (subscript for
`modified Popov')
\begin{eqnarray}
&& \langle \Delta \theta^2( x ) \rangle_{\rm mP} 
= 
\label{eq:mod-Popov-phase}\\
&&
\frac{ 1 }{ \nqc }
\int\!\frac{ {\rm d}k }{ 2\pi }
(1 - \cos k x)
\left\{
\frac{ g \nqc }{ E }
\coth \frac{ \beta E }{ 2 }
-
\frac{ g \nqc }{ \epsilon + \mu }
\right\}
\nonumber
\end{eqnarray}
The first term in curly brackets is proportional to the
anomalous average of Bogoliubov theory~(\ref{eq:Bogo-anomalous}),
the second one is the same counterterm as in the
thermal density $n'$ [Eq.(\ref{eq:mod-Popov-nprime})] and
makes the integral converge in the UV. The IR singularity
of the integrand is the same as in Mora-Castin 
theory~(\ref{eq:MC-phase-fluctuations-normal}),
so that at large distances, a similar phase diffusion
is found:
$\langle \Delta \theta^2( x ) \rangle \approx | x | / 
\ell_\theta$ with
$\ell_\theta = \nqc \lambda^2$. 
The phase coherence length hence grows linearly with
the quasi-condensate density. The plots in 
Fig.\ref{fig:g1G2-correlations}(\emph{top}, \emph{center}) 
illustrate that the difference $\nqc < \mean$
makes the predicted phase coherence somewhat
smaller than in extended Bogoliubov theory (thin solid).
Hartree-Fock theory is even less coherent,
as shown in the top panel.

At zero temperature, the phase fluctuations are sub-diffusive
and increase logarithmically ($|x| \gg \xi \approx \xi_q$)
\begin{equation}
T = 0:\quad
\langle \Delta \theta^2( x ) \rangle_{\rm mP} 
\approx
- \frac{ \log( 2 |x| / \xi ) + \gamma - \pi/\sqrt{2} }{ 2\pi \mean \xi }
\,,
\label{eq:mP-zero-T-G1}
\end{equation}
The power law that this implies for $G_1(x)$ 
[Eq.(\ref{eq:G1-modPopov})] has the same exponent
as Eq.(\ref{eq:MC-zero-T-G1}), but a slightly different
prefactor.

Finally, to come to density correlations, we note that 
Eq.(\ref{eq:Bogo-d-fluct}) is also valid in the presence
of a quasi-condensate as long as one assumes that the
fluctuations obey Gaussian statistics. We generalize
slightly the expressions of
Refs.\cite{Andersen2002c,AlKhawaja2002b,Proukakis2006c}
to cover the case $z \ne z'$: 
as explained around Eq.(37) in Ref.\cite{AlKhawaja2002b},
the anomalous averages are removed
from Eq.(\ref{eq:Bogo-d-fluct}), and one gets
[Fig.\ref{fig:g1G2-correlations}(\emph{bottom})]
\begin{equation}
G_2( x ) 
= \mean^2 + 2 \nqc n'( x ) + [n'( x )]^2
\label{eq:G2-mPopov}
\end{equation}
Here, the function $n'( x )$ is given by 
Eq.(\ref{eq:mod-Popov-nprime})
with an additional factor $\cos k x$ 
inserted under the integral. 
As noted in Ref.\cite{Proukakis2006c},
the reduction of density fluctuations, relative to the
ideal gas, provides an alternative interpretation
of the quasi-condensate density:
$G_2(0) = 2 \mean^2 - \nqc^2$. On the other hand, since
$\nqc \le \mean$ by construction, one always has
$G_2(0) \ge \mean^2$, and there is no possibility
for anti-bunching in modified Popov theory
[see Fig.\ref{fig:eqn-state-MC-mP-W}(\emph{right})].

The density correlations can be approximated quite
accurately (dotted lines in Fig.\ref{fig:g1G2-correlations}(\emph{bottom})) by using
\begin{equation}
x \gg \lambda: \qquad
n'( x ) 
\approx
\frac{ \xi_q \, {\rm e}^{ - |x| / \xi_q } }{ \lambda^2 }
+
\frac{ \xi \, {\rm e}^{ - |x| / (\sqrt{2}\,\xi) } }{ \sqrt{ 32 }\,\xi_q^2  }
\label{eq:mP-G2-approximation}
\end{equation}
The first term results in a formula similar to 
Eq.(\ref{eq:MC-G2-approximation}),
but involving the quasi-condensate healing length $\xi_q$.
The second is small at low energies
and arises from the counter term.

\subsubsection{Renormalised interactions}
\label{s:HF-m}

The scattering between two atoms in a dense gas occurs in a
`background field' formed by the other atoms. This leads to
an energy- and density-dependent change in the matrix elements
of the interaction potential~\cite{Proukakis1998a}. For completeness, 
we discuss here the formulas given in Ref.\cite{AlKhawaja2002b},
adapted to our notation. 

As a first example, consider two atoms in the condensate 
that collide at zero temperature. 
The bare interaction constant $g$ is replaced 
by the two-body T-matrix element
[Eq.(7) of Ref.\cite{AlKhawaja2002b}]
\begin{equation}
\frac{ 1 }{ T_{2\rm B}( - 2\mu ) } = \frac{1}{ g } + 
\int\!\frac{ {\rm d}k }{ 2\pi } \frac{ 1 }{ 2 ( \epsilon + \mu ) }
\label{eq:def-T2B}
\end{equation}
where the denominator involves the pair's kinetic energy and 
the change in the condensate energy as two atoms are removed.
This integral evaluates to [see Eq.(\ref{eq:def-Ib})] 
\begin{equation}
T_{2\rm B}( - 2\mu ) = \frac{ g }{ 1 + g / ( \sqrt{ 32 }\, \mu \xi ) }
\label{eq:T2b-result}
\end{equation}
and illustrates that the interactions renormalize to zero as
$\mu \to 0$. The magnitude of this effect is small 
in practical one-dimensional systems because the denominator
involves the small Lieb-Liniger parameter 
$g / (2\mu \xi ) \sim (\nc \xi)^{-1/2}$.

Our second example are the thermal corrections to the
scattering matrix.
Consider for simplicity the  
dilute phase and the many-body effects in Hartree-Fock theory.
The average density is
worked out as in Eq.(\ref{eq:HF-density}), with
the mean-field shift of the chemical potential replaced
by the Hartree-Fock self-energy, $2 g \mean \mapsto \Sigma$.
According to Eq.(29) of Ref.\cite{AlKhawaja2002b},
the renormalised T-matrix is
\begin{equation}
\frac{ 1 }{ T_{\rm MB}( - \Sigma ) } = 
\frac{ 1 }{ g } 
+ 
\int\!\frac{ {\rm d}k }{ 2\pi } 
\frac{ \coth \frac{1}{2}\beta[\epsilon + \Sigma - \mu ] 
}{ 2 \epsilon + \Sigma }
\label{eq:TmB-definition}
\end{equation}
where Eq.(\ref{eq:def-T2B}) has been used.
The equations are closed by the self-consistency relation
$\Sigma = 2 \mean T_{\rm MB}( - \Sigma )$, 
Eq.(28) of Ref.\cite{AlKhawaja2002b}.

\begin{figure}[htb]
\centerline{\includegraphics*[width = 0.9\columnwidth]{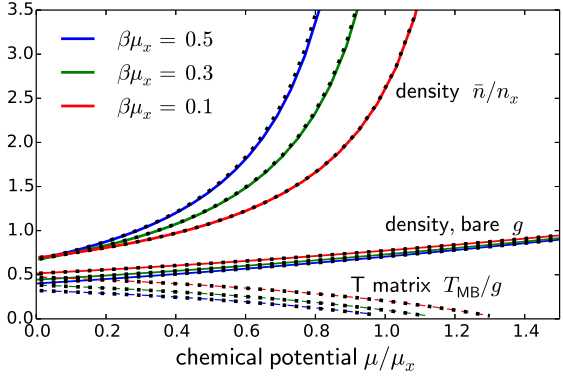}
}
\caption[]{Illustration of renormalized interactions due
to many-body effects for Hartree-Fock theory. Upper set of 
lines: density based on self-consistent 
T-matrix~(\ref{eq:TmB-definition}). Center set: comparison
to density in `bare' Hartree-Fock theory (i.e., constant
coupling $g$). Lower curves: normalized
T-matrix. Dotted curves: low-energy approximations.}
\label{fig:HF-m}
\end{figure}

A numerical solution is shown in Fig.\ref{fig:HF-m} and
illustrates that a critical point appears at 
$\mu \sim \mu_x$ (the precise value depends on $\beta\mu_x$), 
where the many-body interactions renormalize to zero
and the density diverges.
The low-energy approximation 
for Eq.(\ref{eq:TmB-definition}) is [from the techniques of
Appendix~\ref{a:high-T-expansion}]
\begin{eqnarray}
&&
\frac{ 1 }{ T_{\rm MB}( - \Sigma ) } 
\approx
\frac{ 1 }{ g } 
\nonumber\\
&& {}
+ 
\frac{ k_B T }{ \sqrt{ (\Sigma/\mu - 1) \Sigma } }
\frac{ 1/(2 \mu \xi) }{ \sqrt{ \Sigma/ 2 } 
	+
	\sqrt{ \Sigma - \mu } 
	}
+ \frac{ 2 a_2 }{ k_B T \lambda }
\label{eq:TmB-approx}
\end{eqnarray}
This gives, in conjunction with
Eq.(\ref{eq:ideal-gas-eqn-state})
for the density, a relatively accurate picture
(dotted lines in Fig.\ref{fig:HF-m}).
Note the strong ($\approx 50\%$) reduction of interactions
already for $\mu = 0$.
We find in particular that the self-energy 
approaches $\Sigma \to \mu$ at the critical point.

\subsection{Hartree-Fock-Bogoliubov}
\label{s:HFB-Walser}

The Hamiltonian is approximated in this mean-field theory
by the quadratic expression
\begin{eqnarray}
H &\approx& \epsilon_c L + \int\!{\rm d}z\left\{
\frac{ \hbar^2 }{ 2M } 
\frac{ {\rm d}\hat\psi^\dag }{ {\rm d} z }
\frac{ {\rm d}\hat\psi }{ {\rm d} z }
+ \left(
2 g \mean - \mu
\right) \hat\psi^\dag\hat\psi
\right. \nonumber\\
&& \left. {} 
+ \frac{g}{2} \left( 
m^* \hat\psi^2
	+ m (\hat\psi^\dag)^2
\right)
\right\}
\label{eq:Walser-H}
\end{eqnarray}
where $\hat\psi$ is again the fluctuation operator around
a condensate field $\phi$. The first term is the condensate
energy, the first piece under the integral formally
identical to Hartree-Fock theory~[Eq.(eq:HF-Hamiltonian)],
the total density being split into $\mean = |\phi|^2 + n'$.
The last terms involve the anomalous average $m = \phi^2 + m'$ 
that already appeared in 
Bogoliubov theory~[Eq.(\ref{eq:Bogo-anomalous})].
This Hamiltonian is complemented 
by the generalized Gross-Pitaevskii equation for the
condensate field $\phi$
\begin{equation}
- \frac{ \hbar^2 }{ 2M }
\frac{ d^2 \phi }{ dx^2 } 
+
g \left( \mean + n' \right) \phi
+ g m' \phi^* = \mu \phi
\label{eq:}
\end{equation}
A derivation of these equations has been discussed
by Griffin~\cite{Griffin1996} who also uses the name
`Hartree--Fock--Bogoliubov theory'.
Keeping the anomalous average in full goes back to
Girardeau and Arnowitt (see Ref.\cite{Yukalov2008a} for a
discussion in three dimensions).

In a homogeneous system with real $\phi$, 
one finds the equation of state 
\begin{equation}
\mu = g (\mean + n') + g m'
= g |\phi|^2 + g ( 2 n' + m' ) 
\label{eq:Walser-eqn-state}
\end{equation}
The anomalous average $m' < 0$ reduces the chemical potential 
relative to extended Bogoliubov and to modified Popov 
theory~[Eqs.(\ref{eq:mu-MC}, \ref{eq:mu-mP})]. This has
also been interpreted as a many-body-induced
reduction of the particle 
interactions~\cite{Proukakis1998b, Proukakis1998a}.

The expansion of the operator $\hat\psi$ is the same
as in Bogoliubov theory~(\ref{eq:Bogoliubov-trafo}),
but the amplitudes $u = u(k)$, $v = v(k)$ solve the 
modified system
\begin{equation}
\left( \begin{array}{cc} 
\epsilon + 2 g \mean - \mu & g m 
\\ 
g m^* & \epsilon + 2 g \mean - \mu
\end{array}\right)
\left( \begin{array}{c} 
u \\ v^*
\end{array}\right)
= 
\left( \begin{array}{c} 
E u \\ - E v^* 
\end{array}\right)
\label{eq:}
\end{equation}
One gets the dispersion relation (using 
Eq.(\ref{eq:Walser-eqn-state})):
\begin{eqnarray}
E &=& ((\epsilon + 2 g \mean - \mu)^2 - g^2 |m|^2 )^{1/2}
\\
&=& 
\sqrt{
(\epsilon - 2 g m') 
( \epsilon + 2 g |\phi|^2)
}
\label{eq:Walser-dispersion}
\end{eqnarray}
The dispersion relation has the particular feature that 
it shows a
finite gap, $E( k \to 0 ) = 2 g |\phi| \sqrt{ - m' }$. 
Walser argues, in particular
in Ref.\cite{Walser2004}, that the gap is not
in contradiction with the existence of a Goldstone mode
due to the U(1)-symmetry of the original theory. The 
Bogoliubov modes found here are a `convenient quasi-particle basis' 
to describe the thermodynamic equilibrium state of the Bose
gas. 
The finite gap is essential here to regularize
the theory in the infrared.
For the linear response of a perturbation to the gas,
a different calculation is performed that leads, indeed,
to a gapless spectrum of collective excitations.
The key difference is that the perturbation also affects
the condensate phase
which is treated as a dynamical variable, rather than fixed to
a symmetry-broken value~\cite{Walser2004}.

We note that in the self-consistent HFB theory of 
Yukalov and Yukalova~\cite{Yukalov2008a, Yukalov2014},
a gapless dispersion relation for quasi-particles is
constructed by introducing a
second chemical potential (for the non-condensate particles).
Most of this analysis focuses on three dimensions, however.
We do not discuss this variant further here because
when formulas are extrapolated to the one-dimensional setting,
one finds infrared-divergent expressions similar to 
Bogoliubov theory (Sec.\ref{s:Bogoliubov}). 
The dimensional regularization suggested
in Ref.\cite{Yukalov2014} leads to a negative non-condensate
density, similar to extended Bogoliubov theory (Sec.\ref{s:MC}).

\subsubsection{Mean-field densities}

The parameters $n'$, and $m'$ are determined by consistency
from the moments of the fluctuation operator
$\hat\psi$ in the gaussian ensemble defined by 
Eq.(\ref{eq:Walser-H}), for example
$\langle \hat\psi^\dag \hat\psi \rangle = n'$.
This yields again Eq.(\ref{eq:thermal-density}) as in
Bogoliubov theory, but since the expressions for the
amplitudes $u$, $v$ are different, the resulting integral
is regular
\begin{equation}
n' = \int\!\frac{ {\rm d}k }{ 2\pi }
\left\{
\frac{ \epsilon + g(\nc - m') }{ 2 E } 
\coth\frac{ \beta E }{ 2 }
- \frac12
\right\}
\label{eq:Walser-th-density}
\end{equation}
Similarly, for the anomalous average
$\langle \hat\psi \hat\psi \rangle = m'$, 
one finds
\begin{equation}
m' = - 
\int\!\frac{ {\rm d}k }{ 2\pi }
\frac{ g (\nc + m') }{ 2 E } 
\coth\frac{ \beta E }{ 2 }
\label{eq:Walser-anom-density}
\end{equation}
This is an implicit equation since $m'$ also appears
in the mode energies $E$~[Eq.(\ref{eq:Walser-dispersion})].
The $T = 0$ limit has been evaluated in Ref.\cite{Eckart2008}
in terms of elliptic integrals. It has been shown that the
behaviour of the condensate depletion is qualitatively similar
to Eqs.(\ref{eq:MC-zero-T-density}, \ref{eq:mP-eqn-state-T=0}),
except for a logarithmic correction
$\sim \log(\mean \xi) / \xi$ [Eq.(43) of Ref.\cite{Eckart2008}].

\begin{figure}[tbh]
\centerline{\includegraphics*[width = 1.0\columnwidth]{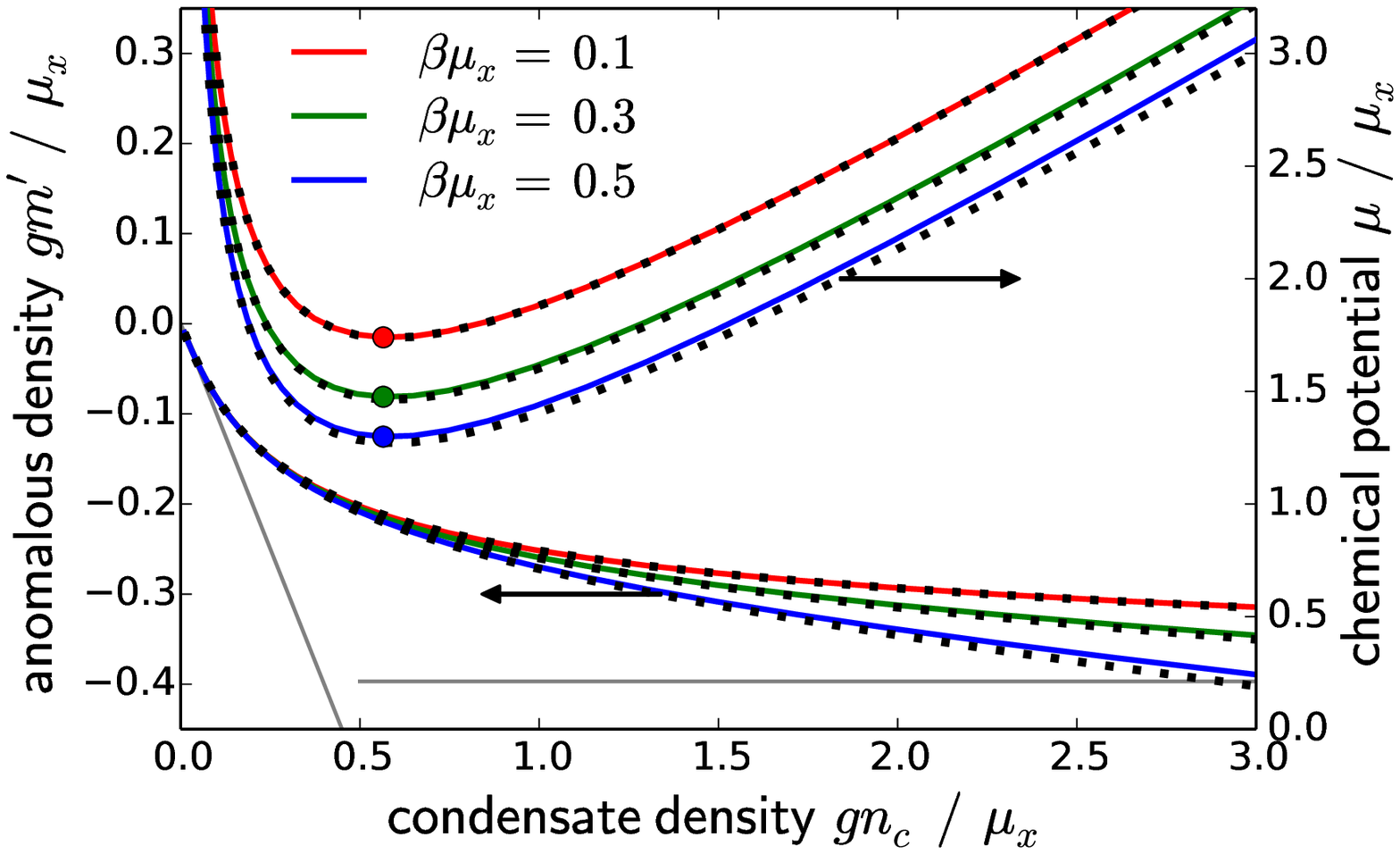}
}
\vspace*{-04mm}

\caption[]{Anomalous average and chemical potential in
Walser's mean field theory (cross-over units).  
\\
Lower set of lines (left scale): anomalous average $m'$,
plotted vs.\ the condensate density. 
{The gray lines correspond
to $m' = - \nc$ and $g m' = -2^{-4/3}\, \mu_x
\approx -0.397\, \mu_x$. 
}
Solid: iterative solution 
based on 
numerical integration of Eq.(\ref{eq:Walser-anom-density});
dotted: low-energy 
approximation~(\ref{eq:mp-low-E-approx_W}).
\\
Upper curves (right scale): 
chemical potential~(\ref{eq:Walser-eqn-state})
based on the self-consistent value for $m'$.
The solid lines are integrated numerically, based on 
Eq.(\ref{eq:Walser-th-density}); the dotted lines
are computed from the 
approximation~(\ref{eq:approx-Walser-th-density}).
The colored dots
mark the `critical point' below which no solution is
found.
}
\label{fig:m-vs-nc}
\end{figure}

In the opposite limit of high temperatures (low
energies, the techniques
sketched in Appendix~\ref{a:high-T-expansion} yield
(correcting one sign in Table~7.1 of Ref.\cite{SauerMSc})
\begin{eqnarray}
n' 
& \approx &
\frac{ \xi_c }{ 2 \lambda^2  }
\left( \sqrt{ \frac{ \nc }{ - m' } } + 1 \right)
-
\frac{ a_1 }{ \lambda }
+
\frac{ a_2 \lambda }{ 4 \xi_c^2 }
\left( 1 + \frac{ m' }{ \nc } \right)
\label{eq:approx-Walser-th-density}
\\
m' &\approx&
- \frac{ \xi_c }{ 2 \lambda^2 } 
\left( \sqrt{ \frac{ \nc }{ - m' } } - 1 \right)
- \frac{ a_2 \lambda }{ 2 \xi_c^2 }
\left( 1 + \frac{ m' }{ \nc } \right)
\label{eq:mp-low-E-approx_W}
\end{eqnarray}
where $\xi_c$
is the healing length corresponding to the condensate 
mean field $g \nc$ and $a_1$, $a_2$ the
Zeta-function coefficients introduced 
earlier~[Eq.(\ref{eq:ideal-gas-eqn-state})].
In Fig.\ref{fig:m-vs-nc}, this approximation is used to
find the anomalous average, yielding the dotted lines.
A coarse estimate for large $\nc$ (horizontal line)
can be found by keeping only the first term in
Eq.(\ref{eq:mp-low-E-approx_W}). For small $\nc$, we note
$m' \approx - \nc$.

The chemical potential $\mu$ is plotted in the same figure
as a function of the condensate density $\nc$.
One notes 
that there is a `critical' chemical potential
$\mu_{\rm cr} = \mu_{\rm cr}( \beta )$
below which the equations have no solution 
[colored dots]. 
To see this intuitively, recall that 
small values of $\nc$ and $m'$ make 
for a large non-condensate density $n'$ 
because of the near-divergence in the infrared
(the first term $\sim n_c^{-1/2}$ 
in Eq.(\ref{eq:approx-Walser-th-density})).
As $\nc$ grows, it eventually dominates in
the chemical potential $\mu = g(\nc + 2 n' + m')$, 
so that the latter
must go through a minimum, which is
located in the range $\mu \sim \mu_x$. 
The same phenomenon occurs in 
modified Popov theory [Fig.\ref{fig:eqn-state-MC-mP-W}],
only the exact location of the `critical point' is different.

\subsubsection{Correlation functions}

The correlation function of the field operator is
\begin{equation}
G_1( x ) = 
\nc + n'( x )
\label{eq:}
\end{equation}
where $n'( x )$ is given by 
Eq.(\ref{eq:Walser-th-density}) with an additional
factor $\cos k x$ under the integral. Due to the gapped
dispersion relation, this is regular in the infrared. It
shows long-range order at the level of the condensate,
$G_1( x \to \infty ) = \nc$
[Fig.\ref{fig:g1G2-correlations}(\emph{center})].
Quite different from the previous theories, this version
of Hartree-Fock-Bogoliubov theory thus predicts a true
condensate (even at one dimension).
The non-condensate contribution has, in the leading order,
the low-energy (and large-distance)
approximation [dotted curve in the Figure]
\begin{equation}
n'( x ) \approx 
\frac{ \xi_c }{ 2 \lambda^2  }
\left( 
\sqrt{ \frac{ \nc }{ - m' } }
{\rm e}^{ - |x| / \xi_m } 
+
{\rm e}^{ - |x| / \xi_c } 
\right)
\label{eq:approx-G1-Walser}
\end{equation}
where $\xi_m \sim (- m')^{-1/2}$ may be called the 
`anomalous healing length' [see Table~\ref{t:length-scales}].

For density fluctuations in Walser's mean field theory, 
we may use
Eq.(\ref{eq:Bogo-d-fluct}) because it is based on a Gaussian
equilibrium ensemble. In distinction to conventional Bogoliubov
theory, all integrals are convergent, and we get for the
pair correlation function
\begin{equation}
G_2( x ) 
= \mean^2 + 2 \nc \left[ n'( x ) + m'( x ) \right]
+ 
[n'( x )]^2 
+ 
[m'( x )]^2 
\label{eq:G2-Walser}
\end{equation}
[for $m'( x )$, insert $\cos k x$
under the integral~(\ref{eq:Walser-anom-density})]. The
analogue of the large-distance 
approximation~(\ref{eq:approx-G1-Walser}) is
\begin{equation}
m'( x ) \approx
- \frac{ \xi_c }{ 2 \lambda^2 } 
\left( \sqrt{ \frac{ \nc }{ - m' } } {\rm e}^{ - |x| / \xi_m }
- {\rm e}^{ - |x| / \xi_c }  \right)
\label{eq:approx-mp-Walser}
\end{equation}
When this is inserted into the density correlation 
function~(\ref{eq:G2-Walser}), it
compares quite well with the numerical calculations,
see Fig.\ref{fig:g1G2-correlations}(\emph{bottom}).

\begin{figure}[tbh]
\centerline{\includegraphics*[width = 0.9\columnwidth]{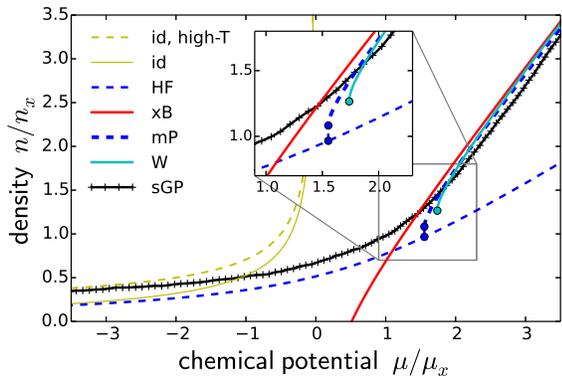}}
\vspace*{-4mm}

\caption[]{{Comparison of equation of state between mean-field
theories (cross-over units).
} 
The temperature is fixed
so that $\beta \mu_x = 0.1$. The inset provides a zoom into
the `critical region'.
Labels:
id, high-$T$: ideal gas, first term in the
low-energy expansion of Eq.(\ref{eq:ideal-gas-eqn-state});
id: ideal gas;
HF: Hartree-Fock;
xB: extended Bogoliubov theory of
Mora and Castin~\cite{Mora2003};
mP: modified Popov theory of Andersen, Al Khawaja et al.~\cite{Andersen2002c, AlKhawaja2002b},
W: Hartree-Fock-Bogoliubov theory,
as developed by Walser~\cite{Griffin1996, Walser2004};
sGP: classical field simulation of the stochastic Gross-Pitaevskii
equation~\cite{ProukakisBook,Cockburn2009, Stoof1999a, Davis2001c, Goral2002}, using the parameters of Table~\ref{t:numbers},
left column.
}
\label{fig:eqn-state-all4}
\end{figure}

\begin{figure}[tbh]
\centerline{\includegraphics*[width = 0.9\columnwidth]{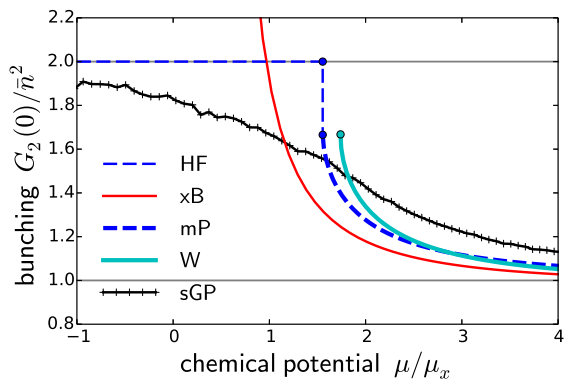}
}
\vspace*{-4mm}

\caption[]{Particle bunching (pair correlations
$G_2(0)$), normalized to the average density squared.
Comparison between mean-field theories and classical field
simulations (black line with markers). Temperature
such that $\beta \mu_x = 0.1$. Labelling as in 
Fig.\ref{fig:eqn-state-all4}.}
\label{fig:G2-all4}
\end{figure}

\section{Discussion}
\label{s:discussion}

\subsection{Complex field simulations and parameters}
\label{s:SGPE}

We use the physical parameters collected in Table~\ref{t:numbers},
left column.
They correspond to a dimensionless
inverse temperature
$\beta \mu_x \approx 0.1$.
In the following plots, we add for comparison 
the results of classical field simulations
(stochastic Gross-Pitaevskii equation~\cite{ProukakisBook})
which were performed for a trapped system 
with these parameters and
the shallow trap frequency 
$1.4\,{\rm Hz} \approx 0.1\,\mu_x/h$.
A real-space
grid with spacing $\Delta z \approx 0.5\,\lambda$ was used,
slightly larger than the `canonical' choice
in extended Bogoliubov theory,
$\Delta z = \xi \beta\mu$
[Eq.(176) of Ref.\cite{Mora2003}].
{If we estimate with Hartree-Fock theory the density
of atoms at energies above the cutoff $E_{\rm max} 
\sim \hbar^2 /(M \Delta z^2)$ (not captured by the simulations), 
we find a negligible contribution for $\mu \sim 0$.}
We plot the data against 
the local chemical potential $\mu - V(z)$, assuming the 
local-density approximation (LDA) is valid. A discussion
of this assumption is provided in
Appendix~\ref{a:beyond-LDA}.

\subsection{Equation of state}

In Fig.\ref{fig:eqn-state-all4}, we compare the equations
of state for all mean-field theories discussed so far.
Coming from negative $\mu$, Hartree-Fock allows to enter
smoothly the cross-over region. It fails by $50\%$ in
the dense phase, however. One might switch to extended 
Bogoliubov theory (xB) at $\mu \sim \mu_x$ where
the two equations of state cross, but this prescription
is lacking a more detailed justification. The self-consistent
modified Popov theory involves a finite jump in the density
when one switches from HF to its `end point' 
(inset of Fig.\ref{fig:eqn-state-all4}). A similar
jump appears when the HFB theory proposed by Walser
is taken. In the dense phase, the three mean-field theories 
converge fairly
well, but the xB density is systematically higher
[see also Fig.\ref{fig:eqn-state-MC-mP-W}(\emph{left})].
Note that the stochastic simulation is able to describe the
density smoothly throughout the cross-over. Its only 
deficiency appears in the dilute phase where it joins
the classical (low-energy or Rayleigh-Jeans) approximation 
instead of the full (Bose-Einstein) prediction of the ideal
gas.

\subsection{Density fluctuations}

A survey of the predictions for density fluctuations is
given in Fig.\ref{fig:G2-all4}. We plot the normalized
pair correlation function $g_2 = G_2(0)/\mean^2$. The ideal
gas and Hartree-Fock theory give a value of $2$ typical for 
a complex Gaussian (or chaotic) field (Wick theorem). 
In modified Popov theory, this jumps at
a `critical chemical potential' $\mu \sim 1.5\,\mu_x$
down to a value $g_2 \sim 1.6$, and decreases 
for $\mu > 1.5\,\mu_x$ steeply to
the `pure condensate' value $g_2 = 1$. From its
critical point on (which is slightly shifted),
Walser's HFB theory behaves similarly. In the extended
Bogoliubov theory, the density fluctuations
diverge as $\mu \to 0$, and one clearly leaves its region of validity.
The stochastic simulation behaves again smoothly and shows
that the suppression of density fluctuations is already
significant
on the dilute side of the cross-over ($\mu < 0$).

\subsection{Conclusion}

We have analyzed the cross-over of a weakly interacting,
homogeneous Bose gas in one dimension in the thermodynamic
limit. Interactions (repulsive) stabilize the dilute phase,
as the chemical potential increases above zero, but at finite
temperature, phase fluctuations persist in the dense phase
and preclude any long-range order (quasi-condensate). 
Using a suitable thermodynamic scaling of 
the relevant variables, the cross-over can be mapped to 
a relatively narrow range of reduced variables, e.g.,
$-\mu_x \lesssim \mu \lesssim 3\,\mu_x$ where 
$\mu_x \sim (g T)^{2/3}$.
We have worked through a portfolio of mean field
theories to describe the cross-over. 
Hartree-Fock theory performs better than 
the ideal gas model, 
but fails to capture the equation of state and the
reduction of density fluctuations in the quasi-condensed 
phase.
This does not seem to improve when the many-body renormalization
of atomic interactions is taken into account.
The extended Bogoliubov theory of Mora and Castin breaks down
when the cross-over is approached from the dense side,
because density fluctuations become too strong. Self-consistent
theories (modified Popov of Stoof et al.,
Hartree-Fock-Bogoliubov of Walser, Holland et al.) predict
a critical point in the equation of state because 
infrared divergences at low (quasi)condensate density enforce
a minimal value for the chemical potential. The failure appears
for both gapped and gapless quasi-particle spectra. 
The issue of constructing a number-conserving theory 
(fixed particle number, canonical ensemble)
is of minor importance for the homogeneous system 
we were focusing on. 
{It can be checked
explicitly from Ref.\cite{Mora2003}
that the specific features (projection of 
quasi-particle modes perpendicular to the condensate, 
condensate phase operator) become irrelevant 
in the thermodynamic limit.
}

We could gauge this state of affairs by comparison to two
successful models for the cross-over. One is provided by the
exact solution of the (Lieb-Liniger) Yang-Yang equations,
{which gives an easy access to low moments of the 
density~\cite{Kheruntsyan2003, vanAmerongen2008, Armijo2010}.
}
The second method builds on complex-field simulations 
(stochastic Gross-Pitaevskii equation) 
that
capture the low-lying modes of the quasi-condensate which can
be described classically. With a suitable choice of numerical
cutoff, these simulations are essentially unique. Their smooth
density profiles through the cross-over region have already compared favorably with experiments. We may expect that 
the distribution functions (counting statistics) that can be 
extracted from them (see, e.g., Ref.\cite{Cockburn2011a}) may
help curing the deficiencies of mean field theories.
{We have reasons to believe that the failures of mean-field
are related to the break-down of
the Gaussian approximation to the probability distribution
of the quantum field.
(For a discussion of beyond-Gaussian correlations in c-field 
methods, see Ref.\cite{Wright2011a}.)
}
This conclusion is based
on the comparison with a classical field theory which will 
be reported elsewhere \cite{Polster20XX}.

\begin{acknowledgments}
T.-O. S. thanks the University of Newcastle and the 
Joint Quantum Centre Durham-Newcastle for their hospitality,
and Vanik E. Mkrtchian and Timo Felbinger for various help
at University of Potsdam.
The stochastic simulations reported here were performed
on the Condor cluster at Newcastle. 
We thank K. Kheruntsyan for sharing results of numerical
calculations of the Yang-Yang model.
We thank Antonio Negretti and Hansj\"org Polster (C.H.) and Stuart Cockburn (N.P) for helpful discussions in various stages of this 
work. Gratefully acknowledged is financial support from the 
\emph{Deutsche Forschungsgemeinschaft} (grant nos.\ Schm-1049/7-1
and Fo 703/2-1, C.H.) and the EPSRC (grant no. EP/F055935/1, N.P.).
\end{acknowledgments}

\appendix 

\section{Low-energy expansions}
\label{a:expansions}

\subsection{Bose function}
\label{a:Bose-g-half}

The Bose function is also known as polylogarithm 
$\mathop{{\rm Li}_{\nu}}( {\rm e}^{x} )$:
\begin{equation}
\mathop{{\rm g}_{\nu}}( x ) = 
\sum_{n=1}^{\infty}\frac{ {\rm e}^{ n x } }{ n^\nu }
= 
\frac{ 1 }{ \Gamma( \nu ) }
\int\limits_0^\infty\!{\rm d}t
\frac{ t^{\nu - 1} }{ {\rm e}^{t - x} - 1 }
\label{eq:def-Bose-fcn}
\end{equation}
The sum converges only for $x < 0$ or a fugacity ${\rm e}^{x} < 1$.
Of interest here is the case $\nu = 1/2$ and the
`high-temperature expansion' 
approaching the critical point from below
\footnote{Eq.(25.12.12) in Digital Library of Mathematical
Functions, \texttt{dlmf.nist.gov}.}
\begin{equation}
x \nearrow 0: \qquad 
\mathop{{\rm g}_{1/2}}( x ) \approx
\sqrt{ \frac{ \pi }{ - x } }
+ \zeta( {\textstyle\frac12} ) + \zeta( -{\textstyle\frac12} ) x 
+ {\cal O}( x^2 )
\label{eq:expansion-Bose}
\end{equation}
with coefficients given by the (analytically continued)
Zeta function.
The first term can be found by expanding the exponential
under the integral~(\ref{eq:def-Bose-fcn}). 
Subtracting this convergent integral and expanding the
integrand for small $x$, we observe that the lowest terms
provide convergent integrals. They yield the following integral 
representations for the $\zeta$-coefficients
\begin{eqnarray}
\int\limits_0^\infty\!\frac{ {\rm d}q }{ \pi }
\left\{
\frac{ 1 }{ {\rm e}^{q^2/2} - 1 }
-
\frac{ 2 }{ q^2 }
\right\}
&=&
\frac{ \zeta( {\textstyle\frac12} ) }{ \sqrt{ 2 \pi } }
= - a_1
\label{eq:a1-integral}
\\
\int\limits_0^\infty\!\frac{ {\rm d}q }{ \pi }
\left\{
\frac{ 1 }{ 4 \sinh^2( q^2/4 ) }
-
\frac{ 4 }{ q^4 }
\right\}
&=&
\frac{ \zeta( -{\textstyle\frac12} ) }{ \sqrt{ 2 \pi } }
= - a_2
\label{eq:a2-integral}
\end{eqnarray}
where the suggestive substitution $t = q^2/2$ was made.
The coefficients are approximately
$a_1 \approx 0.5826$,
$a_2 \approx 0.0830$.

\subsection{High temperature expansion}
\label{a:high-T-expansion}

As an  illustration of the technique, we consider the integral
that appears in the non-condensate density~(\ref{eq:mod-Popov-nprime})
[see also Eq.(\ref{eq:integral-density-Brown})]
\begin{eqnarray}
	I_1( \beta ) &=& \int\!\frac{ {\rm d}k }{ 2 \pi }
	\left\{
	\frac{ ( k^2/2 )
	\coth({\textstyle\frac12}\beta E(k) ) 
	}{ 2 E(k) }
	- {\frac12}
	\right\}
	\label{eq:def-I1}
\\
&\approx&
	\frac{ 1 }{ 2 \beta } - 
	\frac{ a_1 }{ \sqrt{ \beta } }
	- a_2 \sqrt{ \beta }
	\qquad (\beta \to 0)
		\label{eq:expansion-I1}
\end{eqnarray}
To simplify the notation in this Appendix, we use units where the
Bogoliubov dispersion relation is 
$E(k) = |k| \sqrt{ 1 + k^2/4 }$.
The dimensionless inverse temperature is $\beta = M c^2 / k_B T$ 
with the speed of sound $c$.

The integrand is even $k$, and we restrict to $0 \le k < \infty$.
Convergence in the infrared is secured by the 
`coherence factor' $k^2 / (4 E(k))$ in front of
the hyperbolic cotangent. 
The classical (high-temperature) limit of the latter
integrates to the first term in 
Eq.(\ref{eq:expansion-I1})
\begin{equation}
\int\limits_{0}^{\infty}\!\frac{{\rm d}k}{\pi}
\frac{ k^2 / 2 }{ \beta E^2(k) }
=
\frac{ 1 }{ 2 \beta }
\label{eq:I1-class-limit}
\end{equation}
The next order arises when this classical limit is subtracted
from the integrand, 
and the high-energy approximation $E(k) \approx k^2/2$ is
applied:
\begin{equation}
\int\limits_{0}^{\infty}\!\frac{{\rm d}k}{\pi}
\left\{
\frac{ 1 }{ {\rm e}^{ \beta k^2 / 2 } - 1 }
-
\frac{ 1 }{ \beta k^2 / 2 }
\right\}
=
- \frac{ a_1 }{ \sqrt{ \beta } }
	\label{eq:find-zeta-integral}
\label{eq:}
\end{equation}
using the substitution $q = \sqrt{ \beta }\, k$ and the
identity~(\ref{eq:a1-integral}). Note that this also includes
the last term $-1/2$ (`vacuum subtraction' ) from
Eq.(\ref{eq:def-I1}).

When the terms in Eqs.(\ref{eq:I1-class-limit},
\ref{eq:find-zeta-integral}) are subtracted from the
integrand, we get an expression that is still integrable both
at low and high momentum. We perform again the substitution
$q = \sqrt{ \beta }\, k$ and expand (at fixed $q$) for
small $\beta$. The dispersion relation, for example,
becomes $\beta E(k) = \frac12 q (q^2 + 4 \beta)^{1/2} 
\approx \frac12 q^2 + \beta + {\cal O}(\beta^2/q^2)$.
The resulting integral scales like $\beta^{1/2}$ and,
in the leading order, involves the integrand
\begin{eqnarray}
&&
-
\sqrt{\beta}
\frac{ (q^2 - 8) {\rm e}^{ q^2 / 2 }
+ (q^4 + 16)
- (q^2 + 8) {\rm e}^{ -q^2 / 2 }
}{ 4 q^4 \sinh^2( q^2 / 4 ) }
\nonumber\\
&& = 
-
\sqrt{\beta}
\left\{
\frac{ 1 }{ 4 \sinh^2(q^2/4) }
- \frac{ 8 }{ q^4 } 
+ \frac{ \coth(q^2/4) }{ q^2 }
\right\}
\label{eq:find-integral}
\end{eqnarray}
The second form makes the subtractions quite transparent
that regularize the integrand as $q \to 0$.
The first and one half of the second term yield
$a_2 \sqrt{ \beta }$ from Eq.(\ref{eq:a2-integral}).
The remainder is integrated by parts to make the
derivative of the $\coth$ appear, taking care of the
cancelling poles. We again find the integral of
Eq.(\ref{eq:a2-integral}), but with a different prefactor:
$- 2 a_2 \sqrt{ \beta }$. The sum gives the last term
in Eq.(\ref{eq:expansion-I1}). 

The next order in this expansion would be ${\cal O}(\beta^{3/2})$. 
In modified Popov theory, the last term in the
non-condensate density~(\ref{eq:mod-Popov-nprime})
is integrated elementarily. Since it is temperature-independent,
it `slips' between the $a_1$ and $a_2$ terms in
Eq.(\ref{eq:expansion-I1}).

\subsection{Zero-temperature expansion}
\label{a:zero-T-expansion}

The non-condensate density involves two integrals.
The first one is
\begin{equation}
I_a = \int\!\frac{ {\rm d}k }{ 2\pi }
\frac{ \epsilon - E }{ 2 E }
\label{a:q-depletion-1}
\end{equation}
By adopting the units explained after Eq.(\ref{eq:def-I1})),
a dimensional factor $1/(2\xi)$ is pulled out.
Here, $\xi = \hbar (4 M \mu)^{-1/2}$ for extended
Bogoliubov theory and $\xi \mapsto \xi_q$ for modified Popov.
Make the substitution 
$k = 2 \sinh t$
and get
\begin{eqnarray}
&&
\epsilon 
= 2 \sinh^2 t
\,,\quad
E = 2 \sinh |t| \cosh t
\nonumber
\\
&& I_a = 
\frac{ 1 }{ 2 \pi \xi }
\int\limits_{0}^{\infty}\!{\rm d} t\, 
( \sinh t - \cosh t )
= 
-
\frac{ 1 }{ 2 \pi \xi }
\end{eqnarray}
The second piece is the term:
\begin{equation}
I_b = \int\!\frac{ {\rm d}k }{ 2\pi }
\frac{ g \nqc }{ 2 (\epsilon + \mu) }
\label{eq:def-Ib}
\end{equation}
which reduces in our units with
$k = \sqrt{2}\, q$ to
\begin{equation}
I_b = \frac{ g \nqc }{ 2\pi \sqrt{2} \mu \xi }
\int\limits_{0}^{\infty}\!\frac{ {\rm d}q }{ q^2 + 1 }
= \frac{ 1 }{ 2\pi \xi_q }
\frac{ \pi \xi }{ \xi_q \sqrt{8} }
\label{eq:}
\end{equation}
The zero-point energy density of Bogoliubov theory,
Eq.(\ref{eq:Bogo-zero-pt-energy}), is integrated similarly.
After the substitution $k = \xi^{-1} \sinh t$,
\begin{equation}
\epsilon_0 - \epsilon_c = - 
\frac{ \mu }{ 2 \pi \xi }
\int\limits_0^{\infty}\!{\rm d}t\,
{\rm e}^{ - 2 t } \cosh t 
= 
- \frac{ \mu }{ 3 \pi \xi }
\label{eq:}
\end{equation}
which is Eq.(\ref{eq:Bogo-gd-energy}).

Some correlation functions involve the integral 
\begin{equation}
C_1( x ) = \int\!\frac{ {\rm d}k }{ 2\pi }
\left( 1 - \cos k x \right)
\frac{ \mu }{ E(k) }
\label{eq:def-C1-integral}
\end{equation}
Due to the $1/k$ singularity at $k = 0$, it is logarithmically
divergent as $x \to \infty$. We are interested in its asymptotic
behaviour. 
Recall the definition of the cosine integral
(here, $\gamma \approx 0.577$ is the Euler-Mascheroni constant)
\begin{equation}
\mathop{\rm Ci}( x ) = 
\int\limits_{0}^{x}\!{\rm d}q \frac{ \cos q - 1 }{ q } 
+ \log x + \gamma 
\label{eq:}
\end{equation}
and its asymptotic form 
\footnote{Eq.(6.12.3) in Digital Library of Mathematical
Functions, \texttt{dlmf.nist.gov}.}
\begin{equation}
\mathop{\rm Ci}( x ) \approx
\frac{ \sin x }{ x } + {\cal O}(1/x^2)
\qquad (x \to \infty)
\label{eq:ci-asymptote}
\end{equation}
Take some $k_* < \infty$ and subtract in the interval 
$0 \le k \le k_*$ the leading term $1/E(k) \approx 1/k$
\begin{eqnarray}
\int\limits_{0}^{k_*}\!{\rm d}k
\frac{ 1 - \cos k x }{ E(k) }
&=&
\log( k_* x ) + \gamma - \mathop{\rm Ci}( k_* x )
\label{eq:C1-asymptote-1}\\
&&
-
\frac14
\int\limits_{0}^{k_*}\!{\rm d}k
\frac{ k ( 1 - \cos k x ) }{ \sqrt{ 1 + k^2/4 }
( 1 + \sqrt{ 1 + k^2/4 }) }
\nonumber
\end{eqnarray}
For large $x$, the cosine integral $\mathop{\rm Ci}( k_* x )$ 
vanishes [Eq.(\ref{eq:ci-asymptote})], and
by the Riemann-Lebesgue lemma,
the $\cos kx$ can be dropped from the integrand which is
regular. The remaining integral is elementary
\begin{eqnarray}
&&
-
\frac14
\int\limits_{0}^{k_*}\!{\rm d}k
\frac{ k }{ \sqrt{ 1 + k^2/4 }
( 1 + \sqrt{ 1 + k^2/4 }) }
\nonumber\\
&&
=
- \log\frac{1 + \sqrt{1 + k_*^2/4}}{2}
\label{eq:result-root-integral}
\end{eqnarray}
Consider now the limit $k_* \to \infty$. On the lhs of 
Eq.(\ref{eq:C1-asymptote-1}), the
integrand scales $\sim 1/k^2$ and falls off sufficiently fast
so that one gets the definite 
integral over $k = 0 \ldots \infty$. On the rhs, 
the integrated
term~(\ref{eq:result-root-integral}) 
becomes $- \log(k_*/4)$ so that the logarithms partially compensate. 
Re-instating the dimensional prefactor,
we finally get
\begin{equation}
C_1( x ) = \frac{
\log( 2 x / \xi ) + \gamma
}{2 \pi \xi}
\label{eq:expand-C1}
\end{equation}
To check this numerically, we keep $k_*$ finite
and improve the UV-convergence of the integral 
over $k_* \le k < \infty$ by adding and subtracting 
$1/(k^2/2 + 1)$ under the integral. 
The added term can be integrated explicitly.

To get the full expression for the correlation function,
we recall that the integral~(\ref{eq:G1-exponent-MC}) 
also contains the zero-temperature density (depletion). 
In the limit of large $x$, by the Riemann-Lebesgue lemma,
this piece integrates to 
Eq.(\ref{eq:MC-zero-T-density}) in the leading order. 
Combining with Eq.(\ref{eq:expand-C1}), we get
the result~(\ref{eq:MC-zero-T-G1}).

\section{Details on numerics}
\label{a:numerics}

We solve implicit equations either with a bisection or
an iterative scheme, depending on the convergence rate 
and a priori knowledge about the interval where the solution
will be found.
In some cases, an interpolation based on parametrically calculated
datasets is used. Critical points are determined by
minimising the chemical potential as a function of
the relevant parameters (e.g., the condensate density, 
see Fig.\ref{fig:m-vs-nc}).

\section{Validity of the local-density approximation}
\label{a:beyond-LDA}

\begin{figure}
\centerline{\includegraphics*[width = \columnwidth]{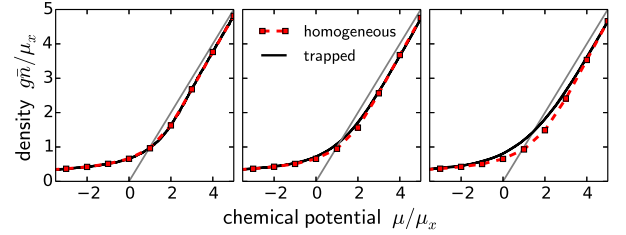}}
\vspace*{-4mm}

\caption[]{
{Comparison of the equation of state for a trapped 
and a homogeneous gas, data from c-field simulations 
with the stochastic Gross-Pitaevskii equation (Sec.~\ref{s:SGPE}).
}
Red dashed with symbols: spatially averaged density in a homogeneous system;
black solid: ensemble-averaged density in a harmonic trap, 
plotted vs.\ the local chemical potential 
$\mu = \mu_t - \frac12 M \omega_z^2 z^2$ and
showing only the border of the (quasi-)condensate. $\mu = 0$
corresponds to the Thomas-Fermi radius, the diagonals give the
lowest-order Bogoliubov result $\mu = g \mean$.
\\
Parameters for Na\,23 atoms given in Table~\ref{t:numbers},
so that $\beta \mu_x = 0.1$.
From left to right, the trap frequency increases through
$\hbar\omega_z \approx 9.94$, $39.8$, $57.2 \times 10^{-3} k_B T$.
(Adapted from Fig.~9.4.2 of Ref.\cite{SauerMSc}.)
}
\label{fig:942-from-Sauer}
\end{figure}

In Fig.\ref{fig:942-from-Sauer}, we compare results obtained
with the stochastic Gross-Pitaevskii equation for trapped
systems with different trap frequencies. 
The red (lower) curve is computed for a homogeneous gas (i.e., a
sufficiently large box with periodic boundary conditions).
The black (upper) curves are based on the local-density
approximation and correspond to 
increasing axial trapping frequency from left to right.
Very good agreement is found on the two asymptotes,
but deviations are visible in the cross-over and grow 
as the trap potential gets steeper. 
This is consistent with the observation that for the
strongest confinement, the inhomogeneity of the potential
is significant on the scale of the cross-over: 
across a displacement of one healing length
$\xi_x$, it changes by a few $\mu_x$.

\vspace*{\fill}


%

\end{document}